\documentstyle[a4,11pt,epsfig]{article}
\newcommand \beq{\begin{eqnarray}}
\newcommand \eeq{\end{eqnarray}}
\def\leftrightarrowfill{$ \mathord\leftarrow \mkern-6mu \cleaders
\hbox{$\mkern-2mu \mathord- \mkern-2mu$}\hfill \mkern-6mu \mathord\rightarrow$}
\def\overleftrightarrow#1{ \vbox{\ialign{##\crcr \leftrightarrowfill\crcr
\noalign{\kern-1pt\nointerlineskip}
$\hfil\displaystyle{#1}\hfil$\crcr}}}
\def\sqr#1#2{{\vcenter{\vbox{\hrule height.#2pt \hbox {\vrule width.#2pt
height#1pt \kern#1pt \vrule width.#2pt} \hrule height.#2pt }}}}

\begin{document}
\title{\sf Neutrino-nucleon scattering rate in proto neutron star 
matter }
\author{\sf L. Mornas$^{(1)}$, A. P\'erez$^{(2)}$  \\ 
\\
{\small{\it (1) Departamento de F{\'\i}sica, Universidad de Oviedo, 
Avda Calvo Sotelo, E-33007 Oviedo (Asturias) Spain}} \\
{\small{\it (2) Departamento de F\'{\i}sica Te\'orica, 
Universidad de Valencia, E-46100 Burjassot (Valencia), Spain }} }
\maketitle
\par\noindent {\small PACS: 26.60.+c, 13.15.+g, 24.10.Jv }
\par\noindent {\small keywords: neutrinos, proto neutron stars, 
relativistic RPA}

\begin{abstract}
We present a calculation of the neutrino-nucleon scattering cross 
section which takes into account the nuclear correlations in
the relativistic random phase approximation. Our approach is based
on a quantum hadrodynamics model with exchange of $\sigma$, $\omega$,
$\pi$, $\rho$ and $\delta$ mesons. In view of applications to neutrino
transport in the final stages of supernova explosion and protoneutron 
star cooling, we study the evolution of the neutrino mean free path
as a function of density, proton-neutron asymmetry and temperature.
Special attention was paid to the issues of renormalization of the
Dirac sea, residual interactions in the tensor channel and meson 
mixing. It is found that RPA corrections, with respect to the mean field
approximation, amount to only 10\% to 15\% at high density.
\end{abstract}

\section{Introduction}

Neutrino transport is a key ingredient to understand the mechanism of 
supernova explosion and the subsequent cooling of the proto neutron 
star formed in the collapse. In the delayed explosion mechanism 
suggested by Wilson, the problem  of shock stall can be overcome 
if sufficient energy is deposited by the neutrinos to revive the 
shock. It is seen in recent numerical simulations 
\cite{BG93,Mezza,Bruenn,RJ00,Yamatr,Ja01} 
of core collapse that, besides other issues such as convection, 
the final outcome of the explosion depends sensitively on the rate 
of neutrino energy deposition.

The following stage of protoneutron star cooling also depends
on the rate of neutrino transport. It determines the shape 
of the neutrino signal emitted during the first few seconds of
the protoneutron star formation. It has also been suggested that
the deleptonization could trigger an instability as the star
cools down, and lead to the delayed formation of a black hole,
in order to explain the time interval observed between the
8th and 9th neutrino detected in the explosion of the
SN1987a  supernova. It is thus important to refine the models, 
since we now begin to benefit of the observation of young neutron star 
surface temperatures, and of the upgrading of neutrino detectors 
such as SuperKamiokande, MACRO, SNO or AMANDA (see e.g. \cite{snews}) 
to collect observational data from an eventual supernova explosion. 

An enhanced neutrino emission can occur at high densities if 
the neutrino-nucleon cross section is modified through the nuclear 
correlations, as was recognized in the early works of Iwamoto and 
Pethick \cite{IP82} and Sawyer \cite{S89}. Interest in this topic 
has recently been revived by the feasability of full Boltzmann 
simulations of neutrino transport in collapsing supernovae 
\cite{Mezza,Bruenn,RJ00,Yamatr}. 

While early calculations only were taking into account the nucleon 
mass reduction in the medium, important progress has been realized 
since then in order to include Hartree-Fock \cite{FabbriMatera}, 
random phase approximation (RPA) 
\cite{HW91,KPH95,RPLP99,LattProc,BS98,NHV99,YT00} or ladder 
\cite{S97,RSS96,RS95,S99,Ya00,Sedra,Voskr} correlations.
It is generally found that the neutrino opacities are suppressed by
medium effects, if however no collective mode is excited \cite{NHV99}.
As a matter of fact, the latter authors find a sizeable enhancement
due to the excitation of a spin zero sound mode, announcing a
transition to a ferromagnetic state in a nonrelativistic treatment.
Due to the degree of uncertainty existing on the equation of state
of neutron star matter at high density, the existence of such
a transition remains an open question, although it seems unfavoured 
in a relativistic treatment. Even if no such instability exists, 
it is important to check the order of magnitude of the suppression 
factor to be applied to the scattering rate. 

In order to asses the importance of some points raised in previous
calculations, we present in this paper a new calculation of the
neutrino-nucleon rate in dense, hot, asymmetric matter. After a
short presentation of the relativistic meson exchange model
and relevant formulae in section two, we first perform a few 
calculations in a simplified model where the background matter
is approximated by pure neutron matter. In particular, we examine 
the effect of the Dirac sea and of various renormalization procedures 
commonly used in the litterature in order to subtract the divergences 
in the vacuum fluctuation term. As a matter of fact, calculations 
reported in \cite{JiangWeiZhou} adscribed a sizeable correction 
due to this effect. In contrast, we found here only a minor effect 
on the size and position of the zero sound mode in the longitudinal 
component, which moreover gets washed out by Landau damping 
and integration on the exchanged momentum and energy. These results 
are presented in section three.

We also studied the way of introducing the short-ranged residual 
interaction in the tensor channel, which has to be taken into 
account in order to avoid the appearance of unobserved pion 
condensation at low density. In the non relativistic formalism, 
this is commonly done by adding a Landau-Migdal contact interaction. 
In the relativistic case, this has been done by various methods, 
one beeing Horowitz's Ansatz in the pion propagator \cite{KPH95,HP94}. 
Two previous works \cite{RPLP99,YT00} using this Ansatz 
found their results to be very sensitive to the choice of the 
numerical value of the Landau-Migdal constant $g'$. We compare 
results obtained with this Ansatz to those obtained from a more 
transparent and straightforward way of introducing this term 
through a contact Lagrangian. We conclude that there is no 
dependence on the value of $g'$, once the contact term is properly 
taken into account, and point out a possible explanation for the 
discrepancy with the authors previously quoted. These results
are presented in section four.

In a second part of this paper, corresponding to section four,
we study the full model where the background matter is 
asymmetric. The proton fraction may be given as a fixed value, or 
be determined by $\beta$ equilibrium in the two extreme situations 
of neutrino free matter $Y_\nu=0$, or trapped neutrinos with 
a lepton fraction typical of supernovae $Y_L\simeq 0.4$.
Here we point out the fact that the phenomenon of meson mixing
has not been taken fully into account in the previous treatments.
Indeed, in addition to the usual $\sigma$-$\omega$ mixing,
the neutron-proton asymmetry also opens the possibility of
$\sigma$-$\rho$ and $\omega$-$\rho$ mixing in the case of 
neutral current relevant for neutrino-nucleon scattering.
Finally, we consider here the contribution of the $\delta$ meson.
This meson is usually left out, since the $\sigma$, $\omega$ and 
$\rho$ mesons are sufficient to describe the properties 
of nuclear matter at the mean field level. In asymmetric matter
however, it can have a non negligible contribution and it mixes 
with the $\sigma$, $\omega$ and $\rho$ mesons.

\section{Neutrino-nucleon scattering in the relativistic 
random phase approximation}

\subsection{Free scattering rate}

Two related processes are needed  in order to calculate the
mean free path and spectrum of neutrinos in matter.
They are the neutrino-nucleon scattering through the neutral
current
\beq
\nu(K) + n(P) \rightarrow \nu(K') + n(P')
\eeq
and the absorption (or creation by the inverse process) through the 
charged current
\beq
\nu(K) +  n(P) \rightarrow  e^-(K') + p(P') 
\eeq  
In either case, the neutrino-nucleon cross section may be written as
\beq
{d \sigma \over d E_\nu d \Omega} = {G_F^2 \over 64 \pi^3} 
{E_{\nu '} \over E_\nu} {\cal I}m (S^{\mu\nu} L_{\mu\nu})
\eeq
\noindent
We will concentrate in this work on the neutral current process.

\par\noindent
Note that we do not include in this definition the Pauli blocking
factor for the outgoing lepton. It will be taken into account
in the final  stage of the calculation when we compute the mean free
path of the neutrino in matter.
In this formula, the tensor $L_{\mu\nu}$ represents the lepton current,
coupled with the weak vertex $\Gamma_{We}^\alpha = \gamma^\alpha 
(1 - \gamma_5)$. For neutral current scattering with massless
neutrinos, we have
\beq
L^{\alpha\beta} &=& {\rm Tr} \left[ (\gamma.K) \Gamma_{We}^\alpha 
(\gamma.K') \Gamma_{We}^\beta \right] \nonumber \\
& =& 8 \left[ 2 K^{\alpha} K^{\beta} + (K.q) g^{\alpha\beta}
-(K^{\alpha} q^{\beta} + q^{\alpha} K^{\beta} ) \mp i 
\epsilon^{\alpha\beta\mu\nu} K_\mu q_\nu \right]
\eeq
\noindent
$K$ and $K'$ are the four momenta of the ingoing and
outgoing leptons. For neutral current scattering
we have $K^2=K'{}^2=0$. The scattering angle is defined
by $\vec K.\vec K {}'= E_\nu E_\nu' \cos \theta$.
$E_{\nu}$ is the ingoing and $E_{\nu '}=E_\nu - \omega$
is the outgoing lepton energy. The energy loss $\omega$ 
is the zero component of momentum exchange 
$q^\mu = K^\mu - K'^\mu$.

The structure function can be related to the imaginary part 
of the retarded polarization
\beq
S^{\mu\nu}(q) &=& \int d^4\, x e^{i q.x}\ < J^\mu(x) J^\nu(0) > \\
& = & {-2  \over 1-e^{-z}} {\cal I}m 
\Pi_R^{\mu\nu}
\eeq
and the differential cross section takes the form
\beq
{d \sigma \over d E_\nu d \Omega} = -{G_F^2 \over 32 \pi^3} 
{E_{\nu '} \over E_\nu} {1 \over 1-e^{-z}}
{\cal I}m (\Pi_R^{\mu\nu} L_{\mu\nu})
\label{cros}
\eeq
The factor $(1-e^{-z})^{-1}$ with 
$z=\beta(\omega - \Delta \mu)$ arises from detailed balance. 
$\Delta \mu$ is the difference between the chemical potential 
of the outgoing and ingoing nucleons. For the neutral current 
process, $\Delta \mu=0$. For the charged current process,
$\Delta \mu=\hat\mu=\mu_n -\mu_p$.
\noindent
At the mean field level, $\Pi_R^{\alpha\beta}$ is the (retarded) 
polarization
\beq
&& {\cal R}e\ \Pi_R^{\alpha\beta}= {\cal R}e\ \Pi_{11}^{\alpha\beta}
\quad, \qquad 
{\cal I}m\ \Pi_R^{\alpha\beta}= \tanh \left( {\beta \omega
\over 2} \right) {\cal I}m\ \Pi_{11}^{\alpha\beta} \nonumber \\
&& \Pi_{11}^{\alpha\beta}=
-i \int d^4\, p\ {\rm Tr} \left[ \Gamma^\alpha G^{11}(p) 
\Gamma^\beta G^{11}(p+q) \right]
\eeq
with $\Gamma^\alpha$ being the weak vertex to the hadronic current
$\Gamma^\alpha= \gamma^\alpha (C_V - C_A \gamma_5)$. For the
neutral current, $C_V^n=-1/2$, $C_A^n=-g_A/2$ for the neutron 
and $C_V^p=1/2 -2 \sin^2 \theta_W$, $C_A=g_A/2$ for the proton,
with $g_A=1.23$ the axial coupling constant and $\sin^2 
\theta_W=0.232$ is the Weinberg angle.
For the charged current we have $C_V=\cos\theta_c$, $C_A=g_A
\cos\theta_c$, with $\cos\theta_c=0.95$ the Cabbibo angle.
$G^{11}(p)$ is the nucleon propagator. At the mean field level, the
nucleon behaves as a quasiparticle with effective mass $M$,
momentum $P$ and chemical potential $\mu$, and the propagator 
can written as the sum of a vacuum and a density dependent term 
(see {\it e.g.} \cite{QHD,Green}) 
\beq
G^{11}(p)&=& (\gamma.P +M) \left\{ {1 \over P^2 - M^2 + i \epsilon}
+ 2 i \pi \delta(P^2-M^2)\left[ \theta(p_0) n(p_0) + \theta(-p_0)
\overline n(p_0) \right] \right\}
\label{propag} \\
{\rm with} && n(p_0) = {1 \over e^{\beta (p_0 -\mu)} +1} \ , \quad
\overline n(p_0) = {1 \over e^{-\beta (p_0 -\mu)} +1}
\eeq

\subsection{RPA corrections}

In the hot and dense medium through which the neutrino is moving 
in the protoneutron star, the nucleon is not free, but correlated
with the other nucleons through the strong interaction.
Several effects contribute, among which are:
\par\noindent {\bf --} The fact that the nucleons can be described in
the medium as quasiparticles with effective masses and chemical
potentials. This represents the mean field approximation. 
This correction has already been taken into account in the
preceding section in the definition of the nucleon propagator.
\par\noindent {\bf --} One can go a step further and include
 the exchange term at the level of Hartree-Fock correlations. 
This is the approach of {\it e.g.} Fabbri and Matera 
\cite{FabbriMatera}. We will not consider this type of 
corrections here. 
\par\noindent {\bf --} Another consequence of the correlations 
is the broadening of the nucleon width, corresponding to the inclusion
of ladder corrections \cite{S97,RSS96,RS95,Ya00}. This is the 
equivalent for scattering to the Landau-Migdal-Pomeranchuk 
effect studied in the case of neutrino Bremsstrahlung
by \cite{Sedra,Voskr,Schaab}.
\par\noindent {\bf --} Finally, one has to take into account the 
RPA correlations. This is the subject of the present work. 

\vskip 0.1cm

The strong interaction may be modelized by relativistic $\sigma$, 
$\omega$, $\pi$ and $\rho$ meson exchange. We will also include
the $\delta$ meson in a later stage of this calculation (see 
section \S \ref{thermodyn}) in order to investigate some specific
effects related to the fact that the matter of the proto neutron star
has a sizeable proton-neutron asymmetry. We will work in a relativistic 
formalism with an interaction Lagrangian of the quantum hadrodynamics
type:
\beq
{\cal L}_{\rm int} &=&
\overline\psi \left(-g_\sigma \sigma +g_\omega \gamma^\mu \omega_\mu
-{f_\pi\over m_\pi} \gamma_5 \gamma^\mu \partial_\mu \vec\pi.\vec\tau 
-g_\delta \vec\delta.\vec\tau  +g_\rho \gamma^\mu 
\vec\rho_\mu.\vec\tau + {f_\rho \over 2 M} \sigma^{\mu\nu} 
\partial_\nu \vec\rho_\mu.\vec\tau \right) \psi \nonumber \\
&&
-{1 \over 3} b m_N (g_\sigma \sigma)^3 
-{1 \over 4} c (g_\sigma \sigma)^4
\eeq
The coupling of the pion is taken in the pseudovector form, 
since this is known to reproduce better the phenomenology of 
nucleon-pion scattering. It will be seen anyway, at the end of the
calculation, that in fact the pion does not contribute directly 
to the neutral current process. 
The non-linear $\sigma$ couplings $\sigma^3$, $\sigma^4$ are 
introduced in order to obtain a better description of the
equation of state with a value of the incompressibility
modulus and effective mass at saturation density in
agreement with the experimental data.

\vskip 0.1cm

RPA correlations can be introduced by substituting the mean field 
polarization in Eq. (\ref{cros}) by the solution of the Dyson equation
\beq
\widetilde\Pi_{WW}^{\mu\nu}=\Pi_{WW}^{\mu\nu} +  
\sum_{a,b=\sigma,\omega,\rho,\pi ...} \Pi_{WS}^{(a)\, \mu\alpha} 
D_{SS\, \alpha\beta}^{(ab)} \widetilde\Pi_{SW}^{(b)\, \beta\nu}
\eeq 
where the index $W$ stands for a vertex with a weak coupling and $S$
for a vertex with a strong coupling. The tilde
$\widetilde{\Pi}$ indicates a resummed polarization. $D_{SS}$ is the
propagator of the mesons $a={\sigma,\ \omega,\ \rho^0, \pi^0}$
for the neutral current process (we will see later that the pion 
does not contribute) or $a=\rho^\pm,\ \pi^\pm$ for the charged 
current process. The first term of the Dyson equation corresponds to 
the mean field approximation taken in the preceding section.

The Dyson equation can be rewritten as
\beq 
\widetilde\Pi_{WW}^{\mu\nu}=\Pi_{WW}^{\mu\nu} +  
\sum_{a,b=\sigma,\omega,\rho,\pi ...} \Pi_{WS}^{(a)\, \mu\alpha} 
\widetilde D_{SS\, \alpha\beta}^{(ab)} \Pi_{SW}^{(b)\, \beta\nu}
\label{Dyson}
\eeq
which provides a solution in terms of the meson propagator
$\widetilde D_{SS}$ dressed in the RPA approximation.

\subsection{Polarizations}
\vskip 3ex   

Polarization insertions were calculated in the relativistic random 
phase approximation. They are given by the loop integral
\beq
\Pi^{AB} =-i \int {d^4 p \over (2 \pi)^4} {\rm Tr} 
\left[ \Gamma^A G(p) \Gamma^B G(p+q) \right]
\label{allthempolariz}
\eeq
$G(p)$ is the nucleon propagator calculated in the Hartree 
approximation, as given in Eq. (\ref{propag}). The $\Gamma^A$
are the vertices of the interaction, which can be either of
the weak type $\Gamma_W= C_V \gamma^\mu + C_A \gamma_5 \gamma^\mu$
or of the strong type $\Gamma_S$ with $S=\{ \sigma,\, \omega,\,
\rho,\, \pi,\, \delta \}$. Formulae are given for the
polarizations in {\it e.g.} \cite{KPH95,H84,KS85,SMS89,RPL98}.

The polarizations can also be obtained by performing a linear response 
analysis on a Hartree ground state. This was the method used
in the following references, to which the reader is referred 
for a full account of the calculation techniques and meson dispersion 
relations. The method was described in \cite{DA85}--\cite{M01b}, 
with results given for the $\pi$ meson in \cite{DA85}, for the 
$\sigma$, and $\omega$ mesons in \cite{DPS89,DP91} and for 
the $\rho$ in \cite{GDP94,MGP98}. 
The notation we are using here can be found in \cite{M01a}, which
also presents a discussion of some results concerning the 
introduction of the contact term. Results concerning 
asymmetric matter are presented in \cite{M01b}. A new
feature, which has been overlooked in previous determinations 
of the RPA corrections \cite{RPLP99,YT00,BS98}, is the mixing 
between mesons of different isospin when the distribution
functions of the protons and neutrons differ. As a consequence,
the Dyson equation (\ref{Dyson}) acquires a more complex
structure. The formulae needed for the calculation of the 
mixed polarizations and meson propagators were given
in \cite{M01b}.

The polarizations calculated in the linear response analysis
coincide with those available in the litterature from the 
Green function formalism at zero temperature and
in symmetric nuclear matter.
At finite temperature, the real parts coincide, but some
precisions are necessary concerning the imaginary parts.
In the real time formalism \cite{Green}, one defines the Green function,
and self energies as 2x2 matrices with the indices labelling 
the two branches of the time contour. The various components
are related to each other. 
In this paper, we chose to work with the retarded polarization. 

Polarizations are generally obtained under the form of a ``density
dependent'' and a ``vacuum fluctuation'' term. This latter part
entails divergences which have to be subtracted by some renormalization
procedure. Regularization of divergences ocurring in vacuum 
polarization may be performed through the inclusion of a counterterm 
lagrangian with couplings adjusted so as to cancel the infinities.
The subtraction is defined up to some residual finite constant, which
is then determined by choosing a renormalization point where the
polarization and its derivatives with respect to the momentum
and effective mass is required to vanish. 

The criteria guiding this choice will have to be reexamined in the 
light of the recent advances concerning effective field theories. 
As a matter of fact, the various possible renormalization schemes 
unfortunately lead to discrepancies observed among the results 
available in the litterature. 
A particularly distressing example was the case of the $\rho$ meson 
discussed in \cite{MGP98}. One cannot ignore the problem, 
since simply eliminating the vacuum by a normal reordering produces 
pathologies in the dispersion relation. The problem roots in the 
fact that we are working in an effective theory, which is non 
renormalizable in the usual QED sense. A renormalization can still 
be performed at a given order. Nevertheless, the parameters of 
the theory, which are adjusted so as to reproduce available experimental 
data, should enforce the symmetries and scalings of the integrated 
degrees of freedom of the underlying more fundamental theory. 
There is some hope that one could solve the problem using
``naturalness'' arguments \cite{Furnstahl}. Fortunately, 
as we will see in section \S \ref{prelim}, our final result has only 
a negligibly weak dependence in the choice of the renormalization 
condition.

The polarization which enters the definition of the differential 
neutrino-nucleon scattering cross section may be decomposed onto 
orthogonal projectors, formed with the vectors and tensor 
available in the problem, {\it i.e.} the metric $g^{\mu\nu}=
{\rm diag}\, (1,-1,-1,-1)$, the hydrodynamic velocity $u^\mu$
and the transferred momentum $q^\mu$. 
\beq
\widetilde\Pi_{WW}^{\mu\nu} = \widetilde\Pi_T\ T^{\mu\nu} + 
\widetilde\Pi_L\ \Lambda^{\mu\nu}  
+ \widetilde\Pi_Q\ Q^{\mu\nu} +i\, \widetilde\Pi_E\ E^{\mu\nu}
\eeq
\beq
{\rm with }\qquad 
\Lambda^{\mu\nu} &=& {\eta^\mu \eta^\nu \over \eta^2} \quad ; \quad
\eta^\mu = u^\mu - {q.u \over q^2} q^\mu \nonumber \\
T^{\mu\nu} &=& g^{\mu\nu} - {\eta^\mu \eta^\nu \over \eta^2} -{q^\mu
q^\nu \over q^2} \nonumber \\
E^{\mu\nu} &=& \epsilon^{\mu\nu\rho\lambda} \eta_\rho q_\lambda \nonumber \\
Q^{\mu\nu} &=& {q^\mu q^\nu \over q^2} 
\label{projtens}
\eeq
The imaginary unit $i$ appearing before the antisymmetric tensor 
$E^{\mu\nu}$ combines with the $i$ present in the corresponding
term of the lepton current, so that the product involves the imaginary
part of the polarization $\Pi_E$. 
The following properties might be useful
\beq
&& T^{\mu\alpha} Q_\alpha^{\ \nu}=T^{\mu\alpha} \Lambda_\alpha^{\ \nu}=
\Lambda^{\mu\alpha} Q_\alpha^{\ \nu}=\Lambda^{\mu\alpha} 
E_\alpha^{\ \nu}=E^{\mu\alpha} Q_\alpha^{\ \nu}=0 \nonumber \\
&& \Lambda^{\mu\alpha} \Lambda_\alpha^{\ \nu} = \Lambda^{\mu\nu} 
\quad , \quad
Q^{\mu\alpha} Q_\alpha^{\ \nu} = Q^{\mu\nu} \quad , \quad
T^{\mu\alpha} T_\alpha^{\ \nu} = T^{\mu\nu} \quad \\
&& E^{\mu\alpha} T_\alpha^{\ \nu} = E^{\mu\nu} \quad , \quad
E^{\mu\alpha} E_\alpha^{\ \nu} = q^2 \eta^2 T^{\mu\nu} \nonumber
\eeq

When calculating the polarization of the mesons with formula
(\ref{allthempolariz}), a zero sound branch appears in the dispersion 
relation of the pion.
Even though it is weaker than in the nonrelativistic formulation,
it remains a spurious effect, since no such effect is observed 
experimentally. When analyzing the structure of the pion potential, 
it is seen that this is related to the short range behavior.
The introduction of the short range effects actually should come out
of a full many-body calculation. Nevertheless, it is a common
practice to implement it by the definition of a residual interaction. 
In the non relativistic formalism, one adds the Landau-Migdal contact 
interaction $V_C = g' (f_\pi/m_\pi)^2 \delta(r) \sigma_1.\sigma_2$.
In the relativistic case, several Ans\"atze have been suggested.
For example, Horowitz {\it el al.} \cite{KPH95,HP94} modify the 
free pion propagator as follows
\beq 
{1 \over q^2-m_\pi^2} \ \longrightarrow \  G_\pi^{(0)\mu\nu}=
{q^\mu q^\nu \over q^2 -m_\pi^2} -g' g^{\mu\nu}
\eeq
and obtain the pion and rho propagators dressed at RPA level by
inverting $G^{-1}=G_0^{-1}-\Pi^{-1}$, {\it i.e.} explicitely
\beq
\left[ \matrix{ G_\rho^{\mu\nu} & G_{\rho\pi}^\nu \cr
                G_{\pi\rho}^\mu & G_\pi^{\mu\nu} \cr} \right]^{-1}
=\left[ \matrix{ G_\rho^{(0)\mu\nu} &  0\cr
                 0 & G_\pi^{(0)\mu\nu} \cr} \right]^{-1}
  - \left[ \matrix{ \Pi_\rho^{\mu\nu} & \Pi_{\rho\pi}^{\mu\nu} \cr
                \Pi_{\pi\rho}^{\mu\nu} & \Pi_\pi^{\mu\nu} \cr} \right]^{-1}
\label{propHor}
\eeq
In these expressions, the polarizations $\Pi_{\pi\rho}^\mu$ and
$\Pi_{\pi}^{\mu\nu}$ are calculated with a pseudo{\it vector} ``pion'' 
vertex $(f_\pi / m_\pi) \gamma_5 \gamma^\mu$. The true pion is obtained 
by projecting over $q_\mu$. 

Another more conventional possibility is to add a contact term 
to the Lagrangian 
\beq
{\cal L} \ni
g' \left( \displaystyle{f_\pi \over m_\pi}\right)^2 \left(\overline\psi 
\gamma_5 \gamma_\mu \psi \right)\left( \overline\psi \gamma_5 \gamma^\mu 
\psi \right)
\eeq
When the dispersion relations of the mesons are calculated using the
method of linear reponse analysis on the Hartree ground state, this 
leads to self consistent equations for the pion and rho polarizations
\cite{M01a} whose solution is given by (\ref{PiPg},\ref{PiRg}).

Both alternatives modify the polarization of the pion and of the 
transverse mode of the $\rho$ meson as follows
\beq
\Pi_\pi  &\longrightarrow & \ {q^2 \Pi_\pi \over q^2  - g '  \Pi_\pi}  
\label{PiPg} \\
\Pi_{\rho T} &\longrightarrow & \Pi_{\rho T} - 
{\eta^2 q^2 g' \Pi_{\rho\pi}^2 \over 1 + g' \Pi_{\pi T}} 
\label{PiRg} \\
& &  {\rm with}\qquad 
 \Pi_{\pi T}=-{1 \over 2} \Pi_\pi^{\mu\nu} T_{\mu\nu}\quad {\rm and}\quad
\eta^2 q^2 \Pi_{\pi\rho}=-{1 \over 2} \Pi_{\pi\rho}^{\mu\nu} E_{\mu\nu}
\nonumber
\eeq

However, the first procedure presents the drawback of introducing additional 
terms in the dispersion relation related to the unphysical components 
of the auxiliary ``pion''. These terms would be responsible for the 
appearance of spurious branches in the spacelike region. 
A more detailed account of this discussion can be found in \cite{M01a}

\section{Preliminary study in pure neutron matter}
\label{prelim}

The RPA correction to the mean field value of the polarization entering
the definition of the neutrino-nucleon differential scattering cross 
section is
\beq
\Delta \Pi_{\rm RPA}^{\mu\nu} &=& \bigl( \matrix{\Pi_{WS}^{(\sigma)\mu} 
& \Pi_{WS}^{(\omega)\mu\alpha} & \Pi_{WS}^{(\rho)\mu\alpha} & 
\Pi_{WS}^{(\pi)\mu} } \bigr)\ \left( \matrix{ G^{(\sigma)} & 
G^{\sigma\omega}_\beta &  0 & 0 \cr
 \noalign{\smallskip}  
G^{(\omega\sigma)}_\alpha & G^{(\omega)}_{\alpha\beta} & 0 & 0 \cr
 \noalign{\smallskip} 
0 & 0 &  G^{(\rho)}_{\alpha\beta} & 0 \cr
 \noalign{\smallskip} 
0 & 0 & 0 & G^{(\pi)} \cr   } \right) \ 
\left( \matrix{\Pi_{SW}^{(\sigma)^\nu} \cr
 \Pi_{SW}^{(\omega)\beta\nu} \cr \Pi_{SW}^{(\rho)\beta\nu} \cr 
\Pi_{SW}^{(\pi)\nu} } \right) \nonumber \\
&=& \Delta\Pi_{RPA}^L \Lambda^{\mu\nu} +  \Delta\Pi_{RPA}^T T^{\mu\nu} 
+ \Delta\Pi_{RPA}^Q Q^{\mu\nu} + i  \Delta\Pi_{RPA}^E E^{\mu\nu} 
\label{Drpanomix}
\eeq 
Introducing the development of the polarizations and propagators
on the projection tensors (\ref{projtens}) we obtain
(where we have dropped the ``WS'' index indicating that the 
polarization loops connect a weak vertex with a strong one)
\beq
 \Delta\Pi_{RPA}^L &=& \Pi_\sigma G_{\sigma} \Pi_\sigma
\eta^2 + 2\ \Pi_\sigma G_{\sigma\omega} \Pi_{\omega\, L} \eta^2  
-\Pi_{\omega\, L} G_{\omega\, L} \Pi_{\omega\, L} 
- \Pi_{\rho\, L} G_{\rho\, L} \Pi_{\rho\, L} \\
 \Delta\Pi_{RPA}^T &=& -\Pi_{\omega\, T} G_{\omega\, T} \Pi_{\omega\, T} 
+\Pi_{\omega\, E} G_{\omega\, T} \Pi_{\omega\, E} \eta^2 q^2 
-\Pi_{\rho\, T} G_{\rho\, T} \Pi_{\rho\, T} 
+\Pi_{\rho\, E} G_{\rho\, T} \Pi_{\rho\, E} \eta^2 q^2 \\
 \Delta\Pi_{RPA}^E &=& -2 \Pi_{\omega\, T} G_{\omega\, T} \Pi_{\omega\, E} 
-2 \Pi_{\rho\, T} G_{\rho\, T} \Pi_{\rho\, E}\\
 \Delta\Pi_{RPA}^Q &=& -\Pi_{\pi} G_\pi \Pi_{\pi}
\eeq
$\Delta \Pi_{\rm RPA}^{\mu\nu}$ must now be contracted with the
tensor of the lepton current $L_{\mu\nu}$. Using the property
\beq
L^{\mu\nu} Q_{\mu \nu}=0
\eeq
we can see that the pion does not contribute to the neutrino-nucleon
scattering.

\subsection{Response functions}
   
The contraction of the lepton current with the polarization
can be expressed by means of three structure functions $R_1(q)$,
$R_2(q)$ and $R_5(q)$ defined as follows \cite{YT00}
\beq
S^{\mu\nu}(q) &=& \int d^4\, x e^{i q.x}\ < J^\mu(x) J^\nu(0) >
\\  
&=& R_1(q) u^\mu u^\nu + R_2(q) (u^\mu u^\nu -g^{\mu\nu}) +
R_3(q) q^\mu q^\nu + R_4 (q^\mu u^\nu + q^\mu u^\nu)
+i R_5(q) \epsilon^{\mu\nu\alpha\beta} \eta_\alpha q_\beta
\nonumber 
\eeq
Using the properties
\beq
&& L^{00}=8 E_\nu (E_\nu - \omega) (\cos \theta +1) \ , \quad
L^{\mu\nu} g_{\mu\nu} = 16 K.q = 16 E_\nu E_\nu' (\cos\theta -1) \\
&& L^{\mu\nu} \Lambda_{\mu\nu} = {-q^2 \over k^2} L^{00} = 
-{q^2 \over k^2} 8 E_\nu E_\nu' ( \cos\theta +1) \\
&& L^{\mu\nu} T_{\mu\nu} = 8 q^2 + {q^2 \over k^2} L^{00} 
= {q^2 \over k^2} \left[ 8 \omega^2 + 8 E_\nu E_\nu' (3 -\cos\theta) 
\right] \\
&& E^{\mu\nu} L_{\mu\nu} = 8 q^2 (E_\nu + E_\nu')
\eeq
we arrive at
\beq
-2 {{\cal I}m \left( L^{\mu\nu} \Pi^R_{\mu\nu} \right) \over  1-e^{-z}}
= 4 E_\nu E_{\nu '} \left[ R_1 (1 + \cos\theta) + R_2 (3-\cos\theta) 
-2 (E_\nu + E_{\nu '}) R_5 (1-\cos\theta) \right]
\eeq
The structure functions are related to the previous polarizations 
$\Pi^{\mu\nu}=\Pi_{\rm MF}^{\mu\nu}+\Delta\Pi_{\rm RPA}^{\mu\nu}$
\par\noindent $=\Pi_L\, \Lambda^{\mu\nu} + \Pi_T\, T^{\mu\nu} +
\Pi_Q\, Q^{\mu\nu} +i\, \Pi_E\, E^{\mu\nu} $ by
\beq
\noalign{\vskip -0.1cm}
R_1 &=& {-2 \over 1-e^{-z}}\ {\cal I}m \left[ -{q^2 \over {\bf q^2}} 
\Pi_L + { w^2 \over  {\bf q^2}} \Pi_T \right] \\
R_2 &=& {2 \over 1-e^{-z}}\  {\cal I}m \left[ \Pi_T \right] \\
R_5 &=& {2 \over 1-e^{-z}}\ {\cal I}m \left[ \Pi_E \right] 
\eeq
\noindent
In the non relativistic limit, $R_1$ and $R_2$ reduce to the density 
and spin density correlation functions respectively. The axial-vector
structure function $R_5$ appears only in a relativistic treatment.

\begin{figure}[htb]
\mbox{%
\parbox{10cm}{\epsfig{file=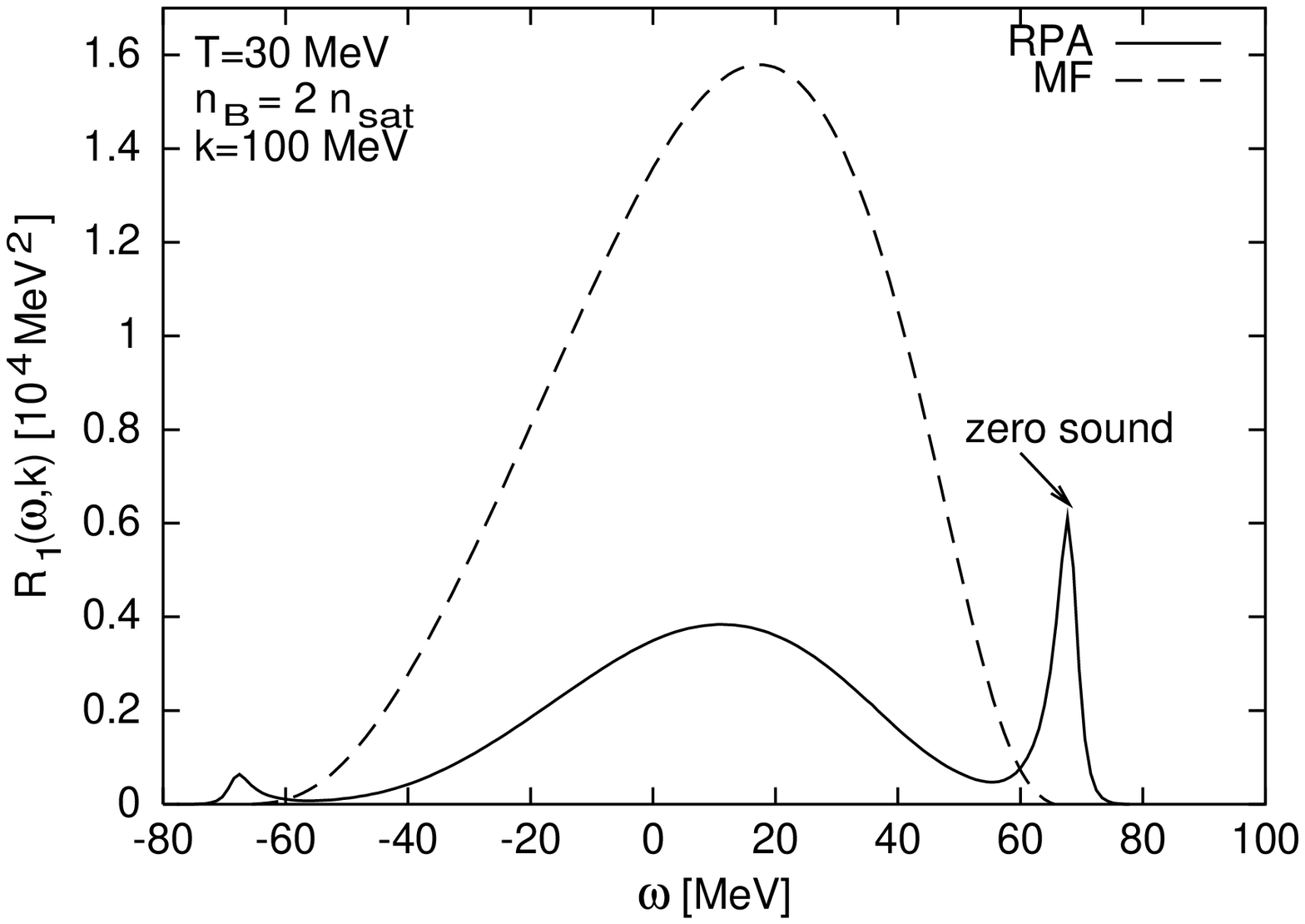,width=10cm}}
\parbox{5cm}{{\bf Fig. 1} -- Structure function R1, displaying overall 
RPA reduction and zero sound enhancement}
}
\end{figure} 

The "vector" structure function $R_1$ involves the sigma meson and
the longitudinal part of the omega meson, as well as the longitudinal 
part of the rho. The contribution of the latter meson is negligible 
in this channel. As is known from previous studies, a zero sound mode 
appears in the mixed $\sigma$-$\omega$ dispersion relation. It manifests
itself as a pole in the RPA meson propagator. At finite temperature, 
it is quenched by Landau damping. As an example, the longitudinal response
function was represented on Fig. 1 at twice the saturation density and
at a temperature $T=30$ MeV for a fixed exchanged 3-momentum $k=100$ MeV.
The longitudinal response function is suppressed in the RPA approximation 
(full line) as compared to the mean field approximation (dashed line).
The zero-sound appears as a peak at the high frequency edge of the
distribution strength. Note, however, that this mode will lie mostly 
outside of the  integration range when we calculate the mean free path. 
Indeed, the fact that the exchanged momentum must be spacelike
restricts the frequency to $\omega^2 < k^2$; moreover, the condition on
the scattering angle $-1 < \cos \theta <1$ imposes that 
$k^2 < (2 E_\nu -\omega)^2$. 
The integration range is therefore restricted to $ k \in [0, \infty [$ 
and $\omega \in [-k, {\rm min}(k, 2 E_\nu -k)]$ or, equivalently, 
$\omega \in [-\infty, E_\nu]$ and $k \in [|\omega|, 2 E_\nu - \omega]$.

\begin{figure}[htb]
\mbox{%
\parbox{10cm}{\epsfig{file=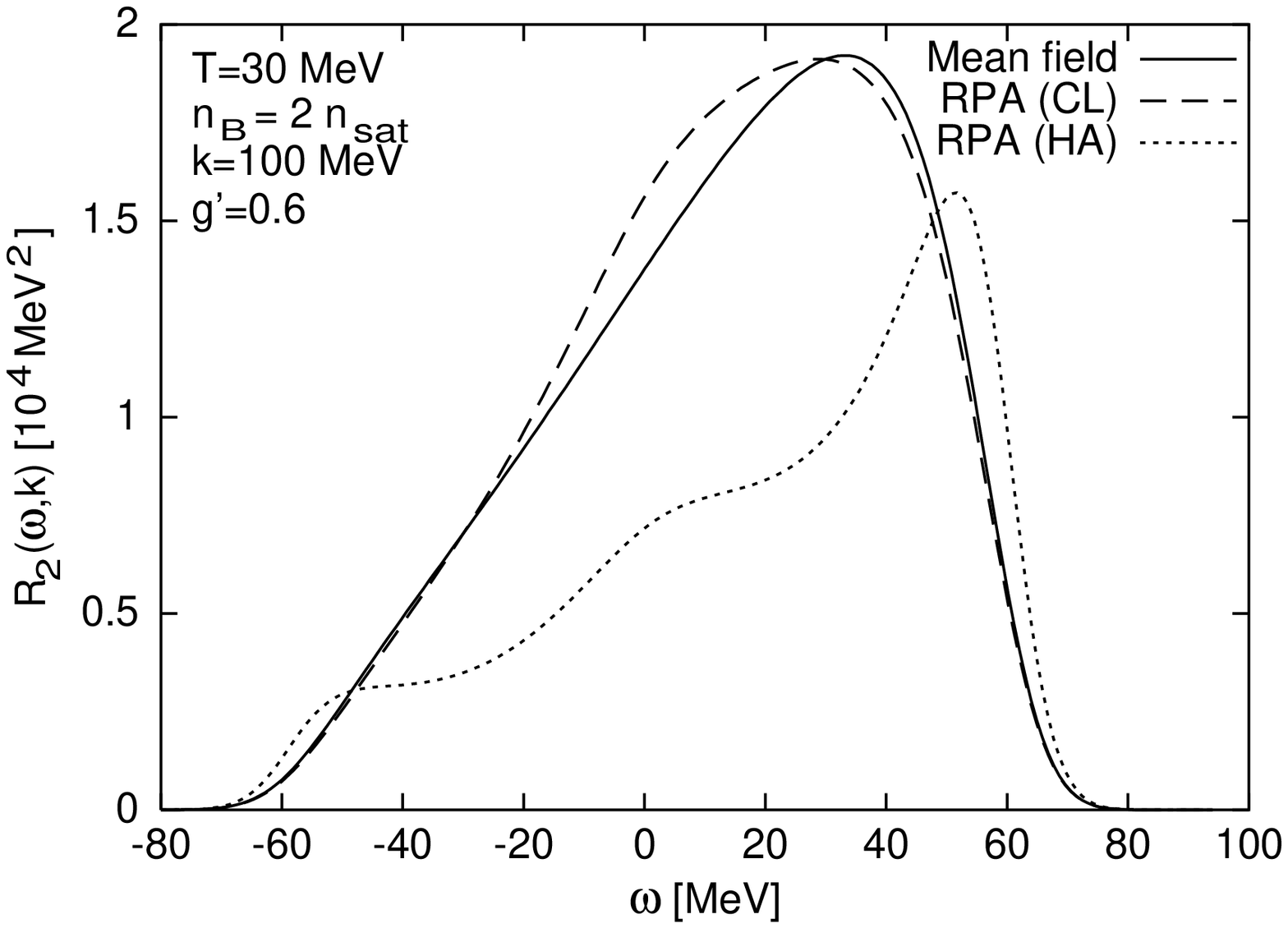,width=10cm}}
\parbox{5cm}{{\bf Fig. 2} -- Structure function $R_2$, with two ways of 
implementing  the short range correlations parametrized by Landau-Migdal 
parameter $g'$}
}
\end{figure} 

The RPA correction to the transverse polarization is mostly 
determined by the $\rho$ meson, whereas the $\sigma$ and $\omega$ 
mesons contribution to this quantity remains small. An example of the 
shape of the transverse structure function is displayed on Fig. 2 
in the mean field approximation (full line) and with RPA corrections 
included (dashed line). The results of Fig. 2 were obtained 
for a density twice that of saturation, a temperature 30 MeV and
at a fixed value of the  3-momentum 100 MeV. 

We find that the dependence in the Landau-Migdal parameter $g'$
is very small, when it is introduced through a contact Lagrangian (CL),
in contrast with the results reported in \cite{RPLP99,YT00}.
The dashed line representing our RPA result is unchanged at the
level of precision of the figure whether we take $g'=0.6$ or $g'=0$.
A tentative explanation for this discrepancy could be the fact
that the later references \cite{RPLP99,YT00} are using Horowitz
Ansatz (HA). Indeed, when the $\pi$-$\rho$ mixings and transversal
projection of the $\Pi_\pi^{\mu\nu}$ tensor introduced through the 
inversion of Eq. (\ref{propHor}) are kept in the equations, 
we would obtain the dotted line of Fig. 2 represented for $g'=0.6$ 
These terms actually would dominate over the other contributions 
from the $\rho$ meson, and would be responsible for a strong suppression 
of the transverse response function. They are proportional to $g'$.
As was argued in the preceding section and in \cite{M01a} however, 
these contributions are spurious and should be subtracted.
This could be the reason for the strong $g'$ dependence observed
by the authors of \cite{RPLP99,YT00}, and of the large reduction 
factors obtained in the RPA approximation in these references. We 
believe that this point needs to be investigated further.  

\begin{figure}[htb]
\mbox{%
\parbox{10cm}{\epsfig{file=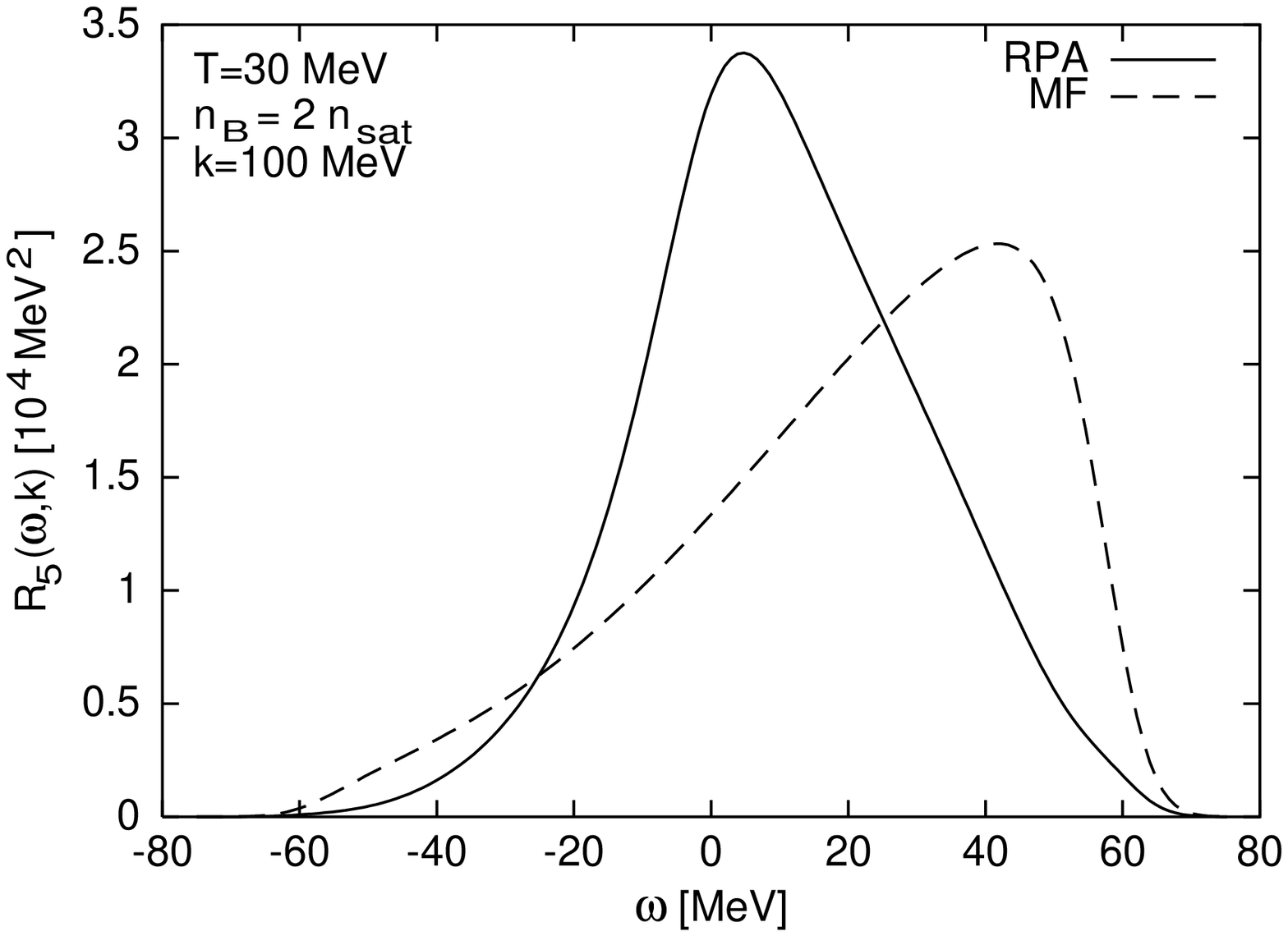,width=10cm}}
\parbox{5cm}{{\bf Fig. 3} -- Structure function R5, displaying a 
shift towards lower frequencies}
}
\end{figure}

As discussed in the previous section, the polarizations entering the
definition of the dressed meson propagators in the RPA approximation
contain a contribution from the vacuum fluctuations which needs to be 
renormalized. At the present time, there exist some discrepancies in 
the litterature concerning the choice of a renormalization procedure. 
As discussed {\it e.g.} in \cite{MGP98,M01a}, this is a source of 
uncertainty in the determination of the dispersion relations of the 
mesons in the framework of hadronic models, and in particular in the 
prediction of the behavior of the effective $\rho$ meson mass in the 
medium. We refer the reader to \cite{MGP98,M01a,M01b,KS88,SH94} for 
details.
Before we proceed, it is necessary to know in what measure does
this issue affect the results presented in this work. Accordingly,
we calculated the response functions in three cases. A first case
consists of simply dropping the vacuum term (which is equivalent 
to perform a normal ordering). In the second case, (``scheme A'' 
in this work) we chose a renormalization scheme as used by Kurasawa 
and Suzuki \cite{KS88} which minimizes the coupling to counterterms 
on the $q^2=0$ shell. In the third case, we used a renormalization
procedure which preserved the formal structure of the expressions of the
vacuum polarizations as a function of the effective nucleon mass
(``scheme 3'' of \cite{MGP98}, called ``scheme B'' in this work).

\begin{figure}[htb]
\mbox{%
\parbox{10cm}{\epsfig{file=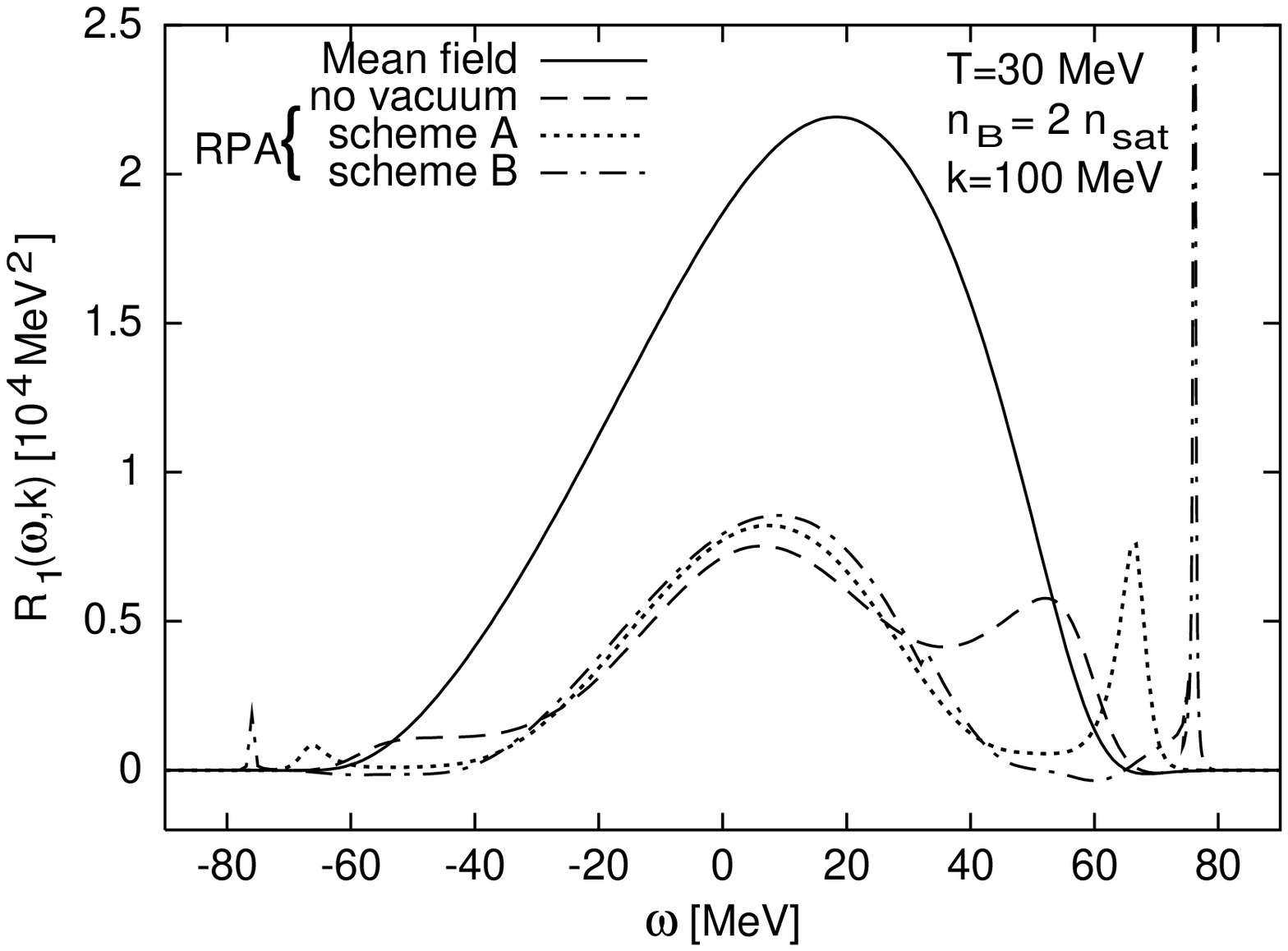,width=10cm}}
\parbox{5cm}{{\bf Fig. 4} -- 
Effect of renormalization scheme on longitudinal 
structure function}
}
\end{figure} 

The choice of the renormalization procedure has some small influence 
on the longitudinal $R_1$ response function. As seen on Fig. 4, its main 
effect is to modify the strength and position of the zero-sound mode.
It was however argued above that the contribution of the zero
sound mode to the neutrino-nucleon scattering is only marginal, since 
its characteristic frequency lies mostly outside of the values
kinematically allowed for neutrino-nucleon scattering. Moreover, 
as will be seen in the next paragraph, the longitudinal reponse
is not the dominant contribution to the neutrino-nucleon scattering.
The structure functions $R_2$ and $R_5$ are not appreciably affected
by the choice of the renormalization scheme. The difference between one or 
the other schemes are of the order of the fraction of a percent and 
therefore not distinguishable on the scale of the figures.
As a consequence, we are able to conclude that the vacuum
fluctuations will not affect the neutrino-nucleon scattering opacity
in any appreciable way.

\subsection{Differential cross section}

The differential cross section can now be calculated from Eq. 
(\ref{cros})
by inserting the previous results for the longitudinal, transverse and
axial vector polarizations. It is instructive to compare the
relative magnitude of their contributions, which are displayed
in Fig. 5. The result is rescaled by dividing it by the textbook 
estimate $\sigma_0= G_F^2 \left( C_V^2 + 3 C_A^2 \right) E_\nu^2/\pi$
\cite{ST83}.
The transverse contribution is dominant and will provide for about
60 \% of the total result. We find that it is very little 
modified by RPA correlations. The corrections arise from the subdominant 
longitudinal and  axial-vector polarizations $\Pi_L$ and $\Pi_E$.
The longitudinal contribution is affected by a reduction factor of 
about two. We notice at the right and left ends of the figure two small 
peaks which arise from the zero-sound mode excitation. 
The contribution of the axial-vector response function is only marginal
at the density ($n_B=2\ n_{\rm sat}$) chosen in this figure, although
it is larger at higher density. It is slightly enhanced by RPA
correlations and its strength is shifted towards lower values
of the energy loss of the neutrino $\omega=E_\nu-E_\nu'$, thus
acting to counterbalance the reduction from the longitudinal
contribution.

\begin{figure}[htb]
\mbox{%
\parbox{10cm}{\epsfig{file=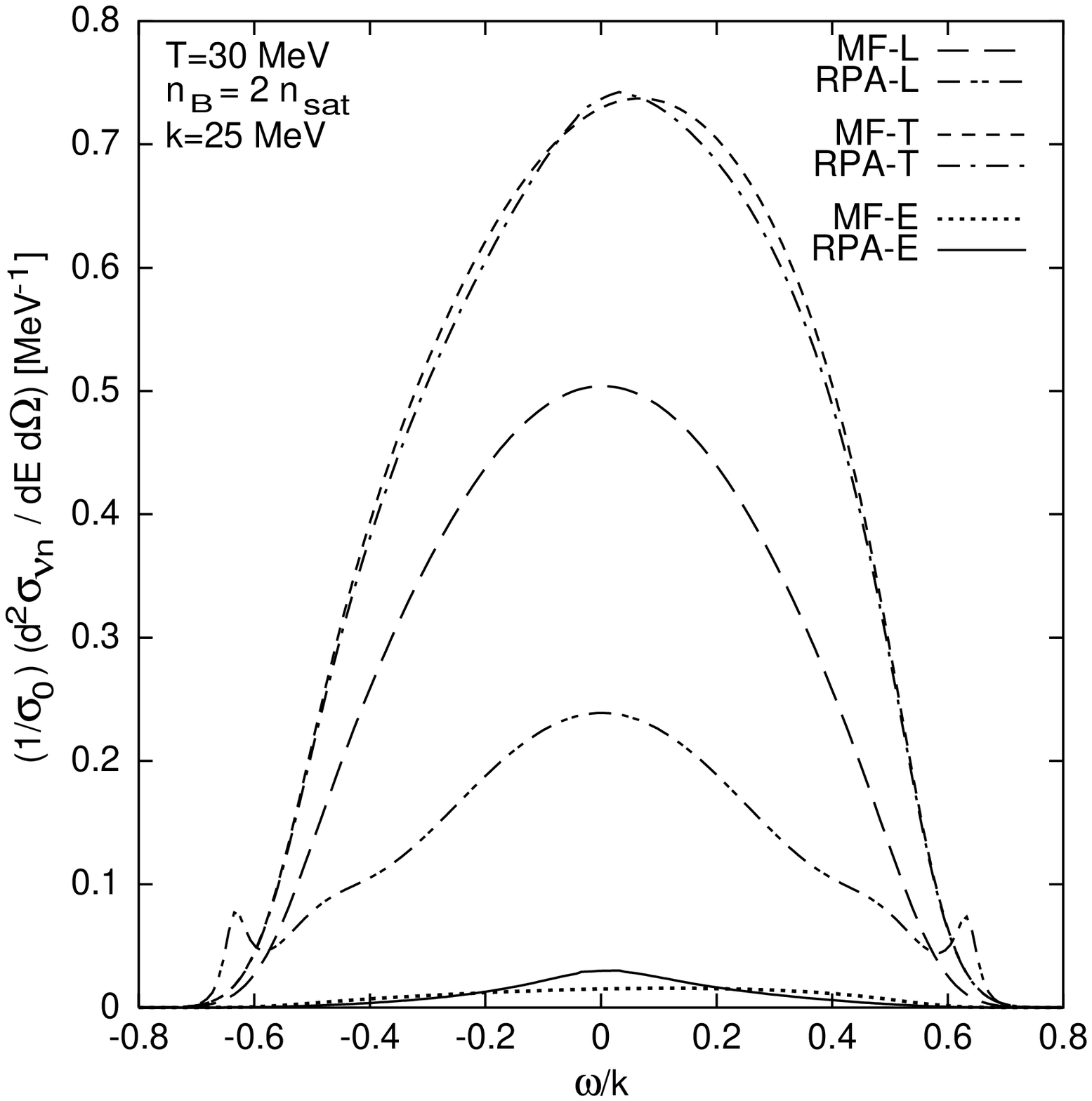,width=10cm}}
\parbox{0.5cm}{\phantom{aaa}}
\parbox{5.5cm}{{\bf Fig. 5} -- Relative contributions of longitudinal, 
transverse and axial vector polarizations to the differential scattering 
cross section}
}
\end{figure} 

As a conclusion to this preliminary study, we gather here the main results 
of the section. We are able to report that the choice of 
renormalization scheme does not affect the results more than on the
level of less than 1 \%. Further good news is that our results
are also insensitive to the choice of the residual contact (Landau-Migdal)
interaction. The transverse response function is left almost 
unchanged by RPA correlations, while the longidudinal and axial
vector responses are reduced by about a factor of two. Since the
transverse structure function yields the dominant contribution to the
neutrino-nucleon scattering, we do not expect spectacular modifications 
from RPA correlations as compared to the mean field result.

\section{Full model - Asymmetric nuclear matter}

Up to now, we have made the simplifying assumption of pure neutron
matter. We now pass to describe a more realistic model, where the
proton fraction is defined by the physical conditions in the star.
The proton fraction is defined by  
\beq
Y_p=\displaystyle{\rho_p \over \rho_p + \rho_n}
\eeq
where $\rho_p$ and $\rho_n$ are the proton and neutron densities.
Numerical calculations show that the chemical equilibrium is
reached very rapidly. The proton fraction will therefore be
determined by $\beta$ equilibrium. As a rule of thumb, in cooled 
neutron stars, the neutrinos can leave the star unhindered, and 
we have $Y_p \simeq 0.1$. On the other hand, in supernovae the 
neutrinos are still trapped dynamically inside the matter on 
the diffusion time scale and contribute to displace the equilibrium
to higher ratios of the proton fraction $Y_p \simeq 0.3$.

The asymmetry enters at several levels in the calculation
of the neutrino nucleon scattering cross section.
It first enters in the determination of the thermodynamics: 
we have $k_{Fp} \not= k_{Fn}$, and also $M_p \not= M_n$  in
a model with $\delta$ meson exchange. It also enters in 
the polarizations, {\it e.g} we must sum the proton and
neutron contributions in polarizations such $\Pi_{\sigma}=
\Pi_{\sigma}^{(nn)}+\Pi_{\sigma}^{(pp)}$. Moreover,
new mixing channels occur which are not present in symmetric 
matter. They arise from polarizations involving vertices with mesons 
of different isospin, which are given by the difference 
between the proton and neutron contributions, such as {\it e.g.}
$\Pi_{\sigma\rho}=\Pi_{\sigma\rho}^{(pp)}-\Pi_{\sigma\rho}^{(nn)}$ 
in longitudinal modes or $\Pi_{\omega\rho}=\Pi_{\omega\rho}^{(pp)}
-\Pi_{\omega\rho}^{(nn)}$ in transverse modes.

\subsection{Thermodynamics}
\label{thermodyn}

The basic properties of nuclear matter can be reproduced 
at the mean field level with $\sigma$, $\omega$ and $\rho$ meson 
exchange, when non linear sigma meson couplings $1/3 b m \sigma^3 
+ 1/4 c \sigma^4$ are taken into account in order to obtain
a better value of the compressibility (see {\it e.g.} \cite{Glend}).

The $\delta$ meson also appears in meson exchange models such 
as the Bonn potential \cite{M87,M89}. Since the $\delta$ meson 
carries isospin, it can give important contributions in 
strongly asymmetric matter.
Moreover, when density dependent mean field models are adjusted 
in order to reproduce the most recent Dirac-Brueckner-Hartree-Fock  
calculations, it is seen that a $\delta$ meson is needed in
order to reproduce the results at finite asymmetry
\cite{Lenske,SST97}.
The behavior of the equation of state when a $\delta$ meson 
is present was studied by Kubis and Kutschera \cite{Kutschera}.
The $\delta$ meson is at the origin of a difference between the
neutron and the proton effective masses. 
\beq
M_n= m -g_\sigma \sigma +g_\delta \delta \quad ,\
M_p= m -g_\sigma \sigma -g_\delta \delta 
\eeq
On the other hand, it brings a negative contribution to the 
asymmetry energy, so that the coupling to the $\rho$ meson has 
to be readjusted to a higher value to compensate for that effect.

In this work, the strength of the $\delta$ coupling $g_\delta$ 
will be fixed to the Bonn potential value \cite{M87} 
$g_\delta^2/4 \pi=1.1075$. 
The couplings $g_\sigma$, $g_\omega$, $g_\rho$, $b$, $c$ were 
adjusted \cite{M01b}
so as to reproduce the experimental saturation point with
\beq
\rho_0 = 0.17\ {\rm fm}^{-3}, \quad B/A=-16\ {\rm MeV}, \quad
K=250\ {\rm MeV}, \quad m_*/m=0.8, \quad a_A=30\ {\rm MeV}
\eeq
and take the values
\beq
&& g_\sigma=8.00, \quad g_\omega=7.667, \quad b=9.637\ 10^{-3}, \quad 
c=7.847\ 10^{-3}, \quad g_\rho=4.59
\eeq
The ratio $\kappa_\rho=f_\rho/g_\rho$ was fixed to the Bonn value 
$\kappa_\rho=6.1$.

The proton fraction is determined by the $\beta$ 
equilibrium condition $\hat\mu = \mu_n - \mu_p = \mu_e - \mu_\nu$.
If neutrinos are trapped inside matter, the chemical potential of 
the neutrino has a finite value $\mu_\nu = (6 \pi^2 \rho Y_\nu)^{1/3}$. 
For a typical value of the lepton fraction $Y_L=Y_\nu + Y_e=0.4$,
$\mu_\nu$ is of the order of 200-250 MeV and the proton fraction 
$Y_p$ of the order of $\sim$ 0.3 - 0.36. $Y_L$ is determined by 
neutrino transport ({\it e.g.}, diffusion equation), to which the 
neutrino-nucleon cross section serves as input. If the matter is 
transparent to neutrinos, $\mu_\nu=0$ and the proton fraction is 
of the order of $\sim$  0.1.
The following figures (Fig. 6) show the behavior of 
thermodynamical parameters
relevant for the calculation of the neutrino opacities. 

\begin{figure}[htb]
\mbox{%
\parbox{7.5cm}{\epsfig{file=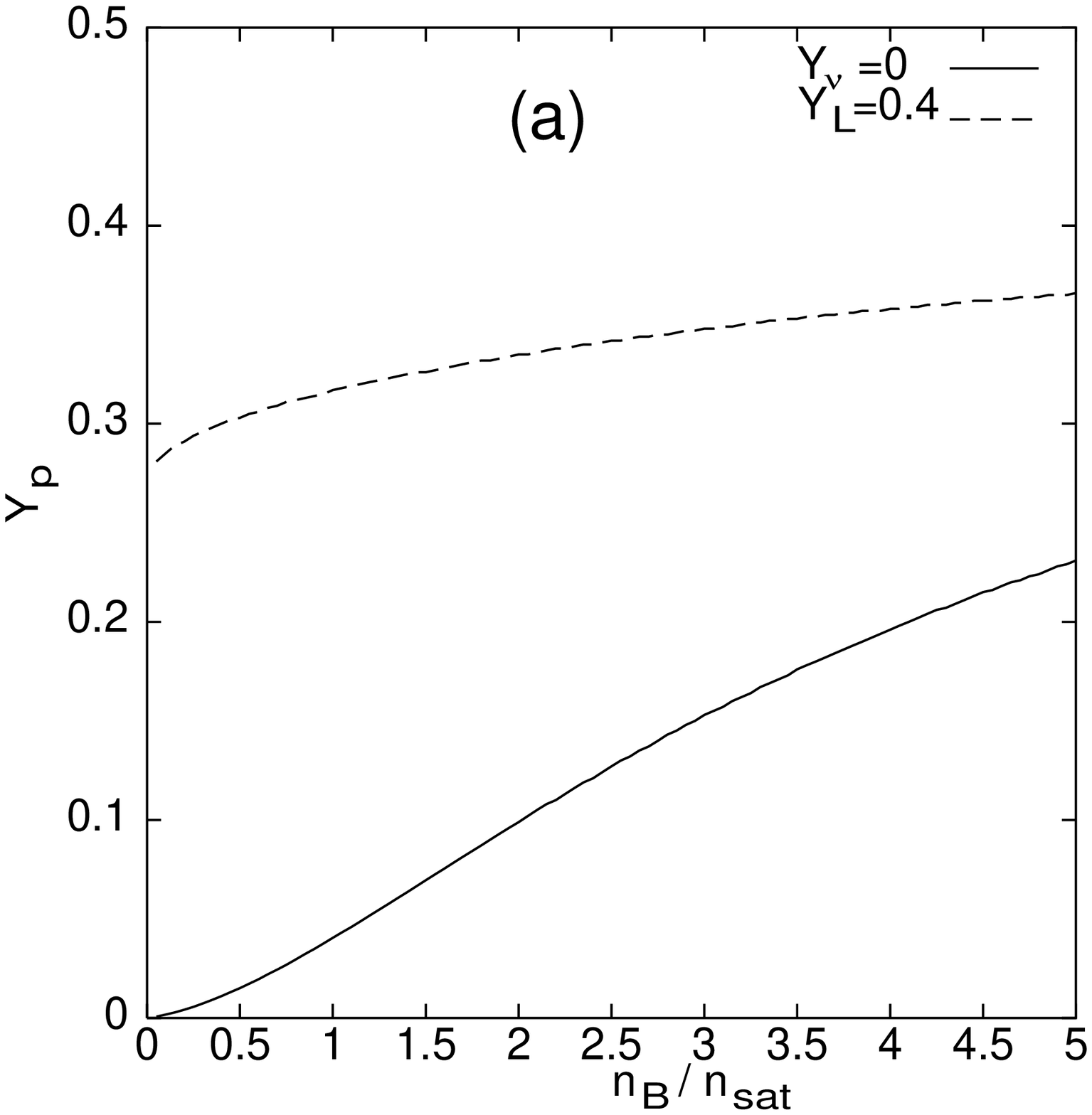,width=7.5cm}}
\parbox{0.5cm}{\phantom{a}}
\parbox{7.5cm}{\epsfig{file=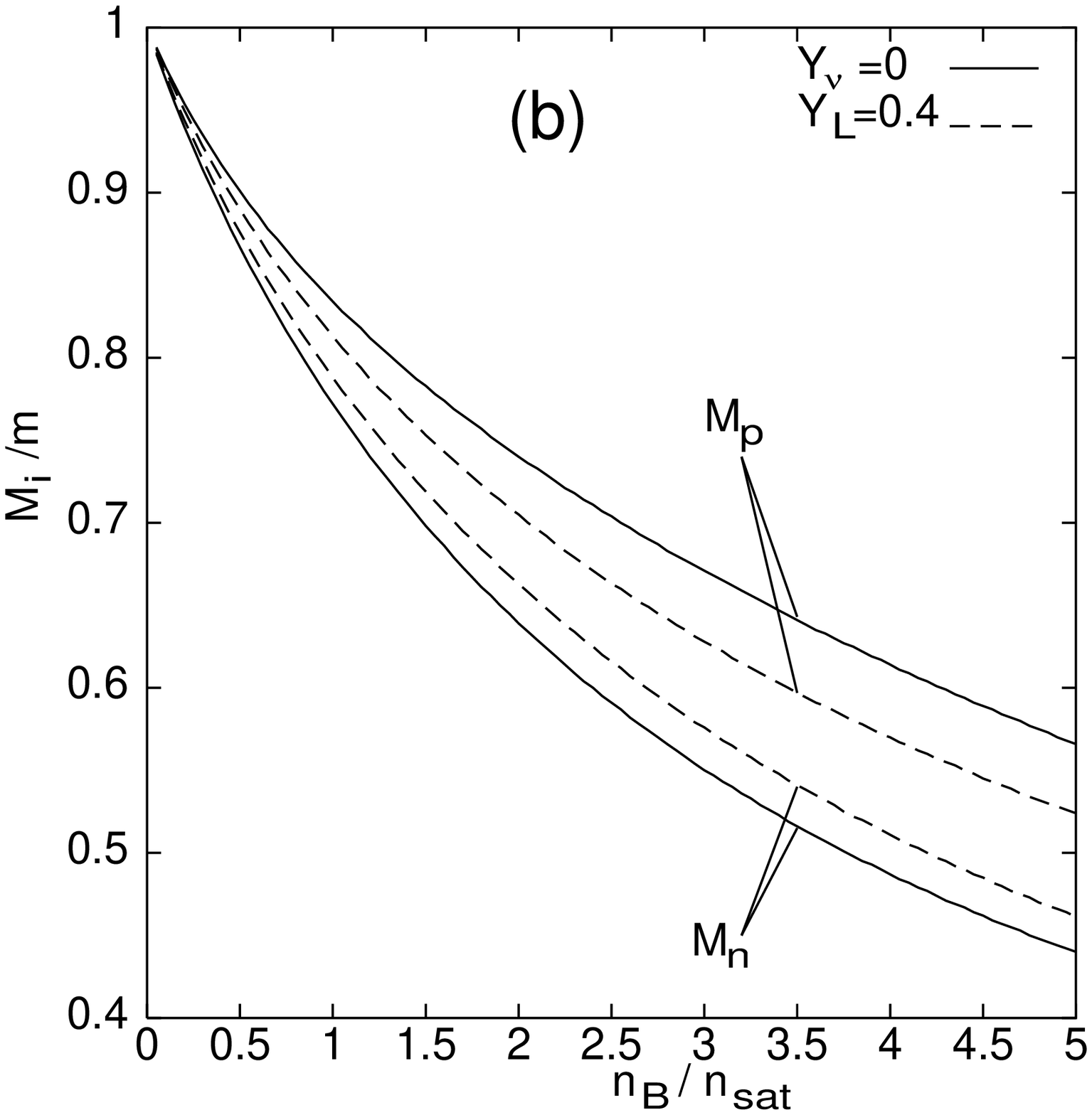,width=7.5cm}}
}
\mbox{%
\parbox{7.5cm}{\epsfig{file=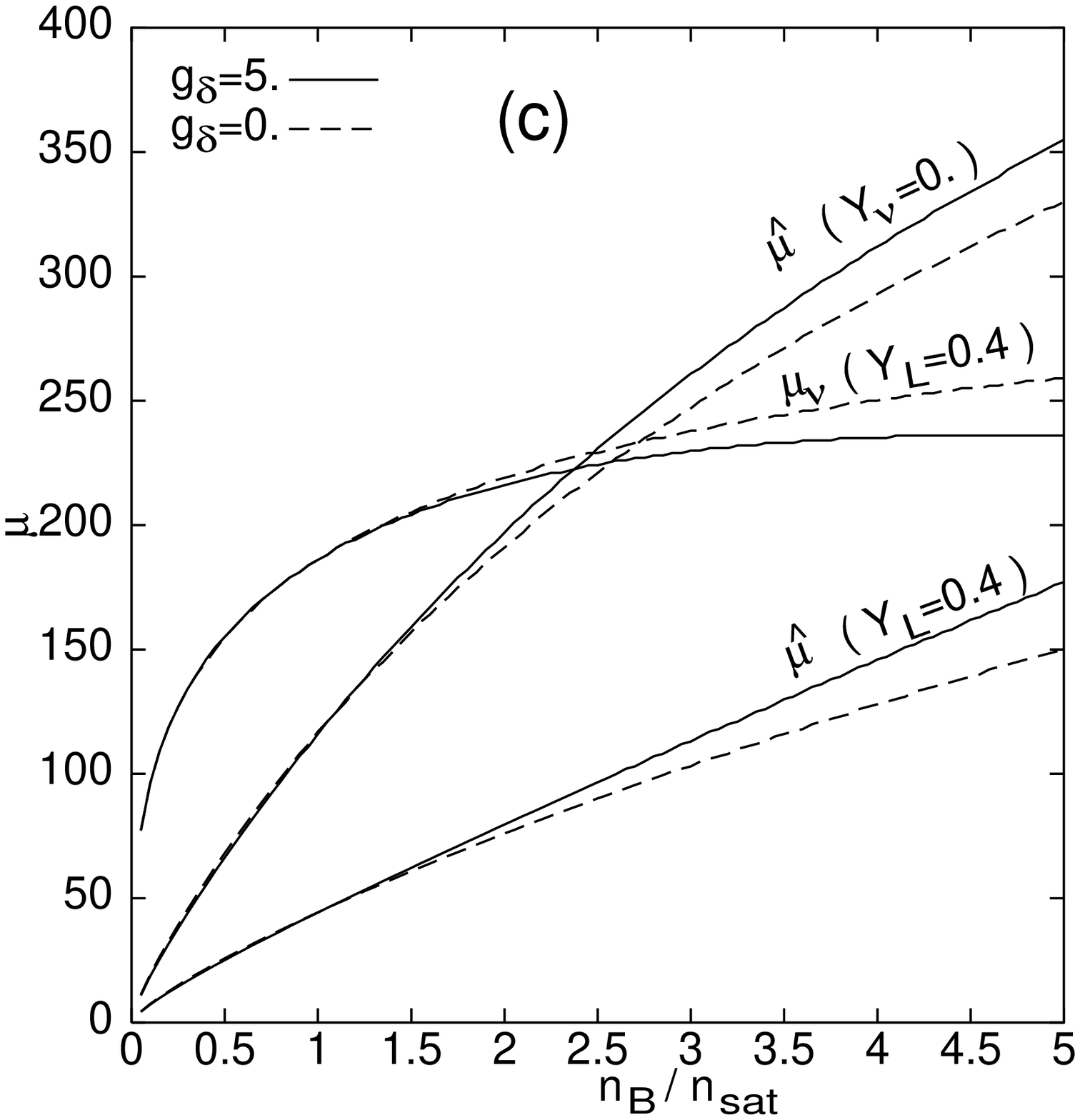,width=7.5cm}}
\parbox{0.5cm}{\phantom{a}}
\parbox{7.5cm}{{\bf Fig. 6} -- Thermodynamical parameters in
neutron star matter in $\beta$ equilibrium, both neutrino free 
($Y_\nu=0$) or with trapped neutrinos ($Y_L=0.4$)}
}
\end{figure} 

The
neutron star matter is assumed to be in $\beta$ equilibrium.
The two cases of neutrino free matter ($Y_\nu=0$) or matter with 
trapped neutrinos ($Y_L=0.4$) are displayed. 
All figures were drawn assuming a vanishing temperature.
The upper left panel shows the evolution of the proton fraction
as a function of the baryon density. The upper right panel
shows the behavior of the proton and neutron effective masses.
The difference $M_p-M_n$ is larger in neutrino free matter
since it allows for a smaller proton fraction and therefore
a larger mean $\delta$ field. On the lower panel are represented
the difference between the chemical potentials of the neutron and 
the proton $\hat\mu$ in the cases  ($Y_\nu=0$) and  ($Y_L=0.4$)
as a full line. In matter with a trapped lepton fraction
$Y_L=0.4$, the neutrino chemical potential is non vanishing;
it is compared to $\hat\mu$ on Fig. 6-(c). The full line
was obtained for a coupling to the $\delta$ field $g_\delta=5$.
If this coupling is set to zero, the chemical potentials 
are given by the dashed lines on Fig. 6-(c).

\subsection{Meson mixing}

The dispersion relations of the $\sigma$-$\omega$-$\delta^0$-$\rho^0$ 
system of neutral mesons, which is relevant for the calculation of 
the neutrino-nucleon scattering through the neutral current process, 
were studied in \cite{M01b}. As discussed in this reference,
the difference existing between the neutron and proton distribution 
functions makes mixing possible in all chanels in asymmetric matter.
The pion does not mix with the other mesons, and, as pointed out
in section \S \ref{prelim}, its contributions to the $\nu$-N 
scattering process vanishes, so we will not consider it further.
The RPA correction is now obtained from
\beq
\Delta \Pi^{RPA} = \left(\Pi_{WS}^{(\sigma)\mu}\quad 
\Pi_{WS}^{(\omega)\mu\alpha} \quad \Pi_{WS}^{(\delta)\mu}\quad 
\Pi_{WS}^{(\rho)\mu\alpha} \right) \times
\left( \matrix{ G^\sigma & G^{\sigma\omega}_\beta & G^{\sigma\delta}
& G^{\sigma\rho}_\beta \cr
\noalign{\smallskip}
G^{\omega\sigma}_\alpha & G^{\omega\omega}_{\alpha\beta} & 
G^{\omega\delta}_\alpha & G^{\omega\rho}_{\alpha\beta} \cr
\noalign{\smallskip}
G^{\delta\sigma} & G^{\delta\omega}_\beta & G^{\delta\delta} & 
G^{\delta\rho}_\beta \cr
\noalign{\smallskip}
G^{\rho\sigma}_\alpha & G^{\rho\omega}_{\alpha\beta} & G^{\rho\delta}_\alpha &
G^{\rho\rho}_{\alpha\beta} \cr } \right) \times \left( \matrix{
\Pi_{SW}^{(\sigma)\nu} \cr \noalign{\smallskip} \Pi_{SW}^{(\omega)\beta\nu} 
\cr \noalign{\smallskip} \Pi_{SW}^{(\delta)\nu}\cr \noalign{\smallskip}
\Pi_{SW}^{(\rho)\beta\nu} \cr} \right)
\eeq
Explicit expressions for the dressed meson propagator matrix were 
given in \cite{M01b}. The calculation proceeds as in section 
\S \ref{prelim}. The longitudinal projection of the RPA 
correction to the polarization now receives contributions from 
the $\delta$ meson and from all possible combinations of 
the mixings, that is, not only from $\sigma$-$\omega$ and 
$\delta$-$\rho$ already present in symmetric matter, but also 
from  $\delta$-$\sigma$, $\sigma$-$\rho$, $\omega$-$\rho$ and
$\omega$-$\delta$ mixings. The transversal $T$ and axial 
vector $E$ components receive additional contributions from
$\omega$-$\rho$ mixing. Finally, the projection parallel to 
$q^\mu q^\nu$ vanishes through contraction with the (massless) 
lepton current, it would contain the contribution of the pion.

\begin{figure}[htb]
\mbox{%
\parbox{8cm}{\epsfig{file=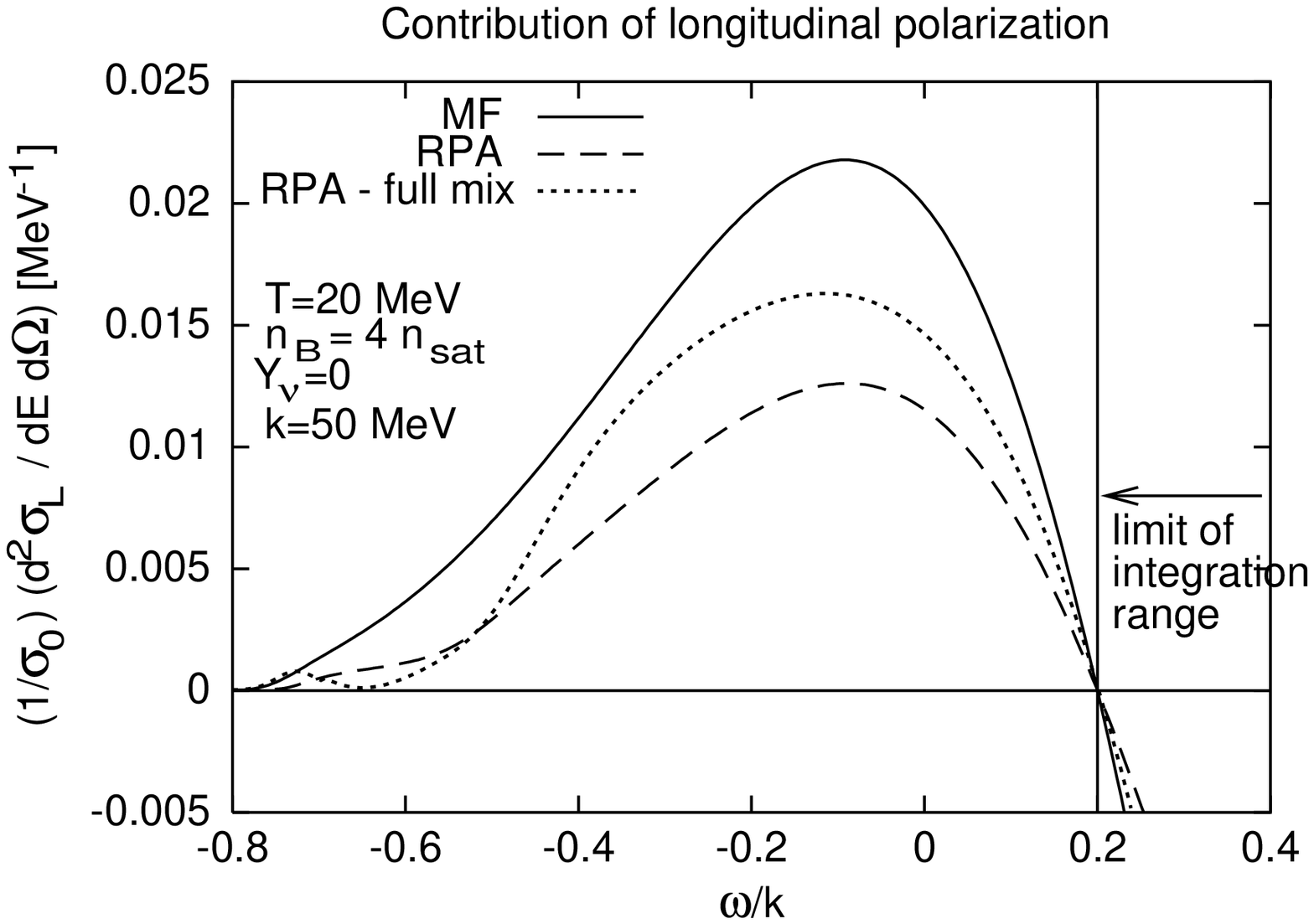,width=8cm}}
\parbox{8cm}{\epsfig{file=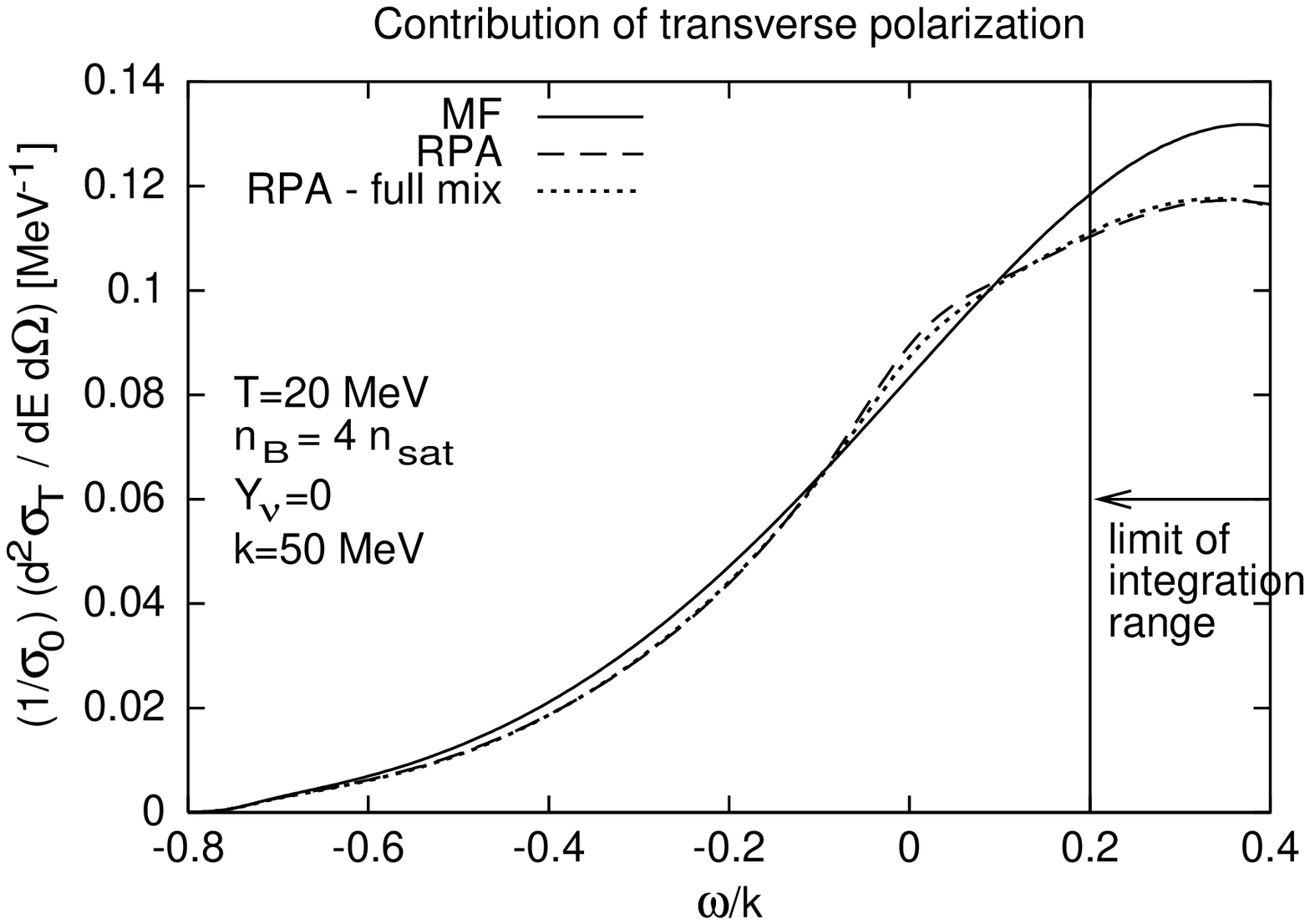,width=8cm}}
}
\mbox{%
\parbox{8cm}{\epsfig{file=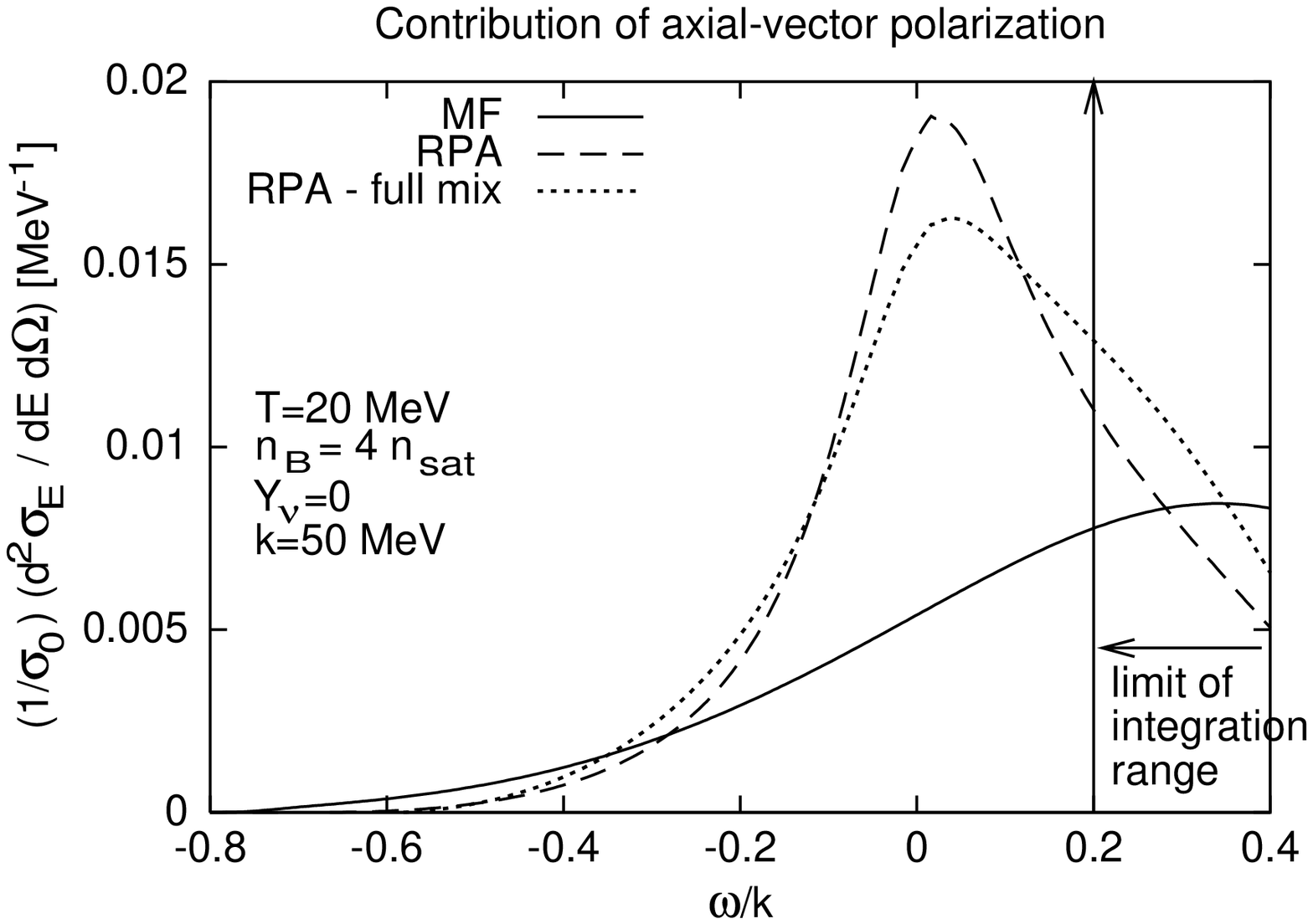,width=8cm}}
\parbox{1cm}{\phantom{aa}}
\parbox{7cm}{{\bf Fig. 7} -- 
Contribution of longitudinal (upper left panel),
transverse (upper right panel) and axial-vector (lower panel) 
polarizations to the differential scattering cross section, 
showing  the correction arising from fully taking into account 
meson mixing.}
}
\end{figure}

The expressions of ${\cal I}m\ ({\Pi_R}^{\mu\nu} L_{\mu\nu})$
so obtained are introduced in the formula for the 
differential cross section (\ref{cros}).
We calculated separately the longitudinal, transverse and 
axial-vector contributions to the differential cross section 
and represent them in Fig. 7 as a function of the ratio
of the neutrino loss energy to the transferred three
momentum $k=50$ MeV. The thermodynamical conditions were chosen
to be those of neutrino free matter ($Y_\nu=0$) with 
a density $n=4\, n_{\rm sat}$ and finite temperature
$T$=20 MeV. The transferred 4-momentum is subject to
kinematical constraints which restrict the range over
which it will be integrated in order to obtain the
total cross section and the mean free path (see next section). 
The limit of the integration range is also represented on
Fig. 7.

It is seen from these figures that the contribution of
the longitudinal polarization is appreciably affected by 
meson mixing. The longitudinal modes involve the full 
$\sigma-\omega-\delta-\rho$ mixing. The mixing is stronger at 
high asymmetry, high density and high momentum transfer.
There exists as before a zero sound mode, but it is undiscernible 
on the figures, because it is partially washed out by Landau damping
and lies outside of the integration range.
 
The transverse and axial vector contributions involve only 
$\rho-\omega$ mixing in the transverse modes. The modification 
to the transverse contribution is small in the kinematically 
allowed region. On the other hand the axial vector response 
function is somewhat enhanced by $\rho$-$\omega$ mixing.

The conclusions that were already presented from the preliminary 
study are confirmed in this more elaborate model. We obtain
that the dominant contribution to the neutrino-nucleon
scattering cross section, that is the transverse one, is hardly
modified by the RPA corrections as compared to the mean field
approximation. The RPA corrections affect the longitudinal 
and axial vector components, the longitudinal part beeing suppressed
and the axial part enhanced with respect to the mean field
approximation, and partially cancelling each other. 
The effect of meson mixing is of the order of 20 \% on both the
longitudinal and axial vector parts, but further act
counterbalance each other. The net result is that we do not
expect dramatic corrections to the global result from
RPA correlations.

\subsection{ Mean free path}

Finally, we will calculate the contribution of neutrino-nucleon 
scattering to the mean free path, since it gives an immediate
feeling of the phenomenon of opacity when it is compared 
with the length scales typical of the star, {\it e.g.} the star 
radius or a typical convection length scale. We must 
naturally stress the fact that this is only an estimate. As
a matter a fact, several other processes, like neutrino 
emision and absorption through charged current, scattering 
on electrons, Bremsstrahlung ... contribute to the definition
of the actual mean free path.

The mean free path is defined here as the inverse of the total 
cross section per unit volume obtained by integrating the 
differential cross section calculated in the preceding sections.
\beq
{1 \over \lambda(E_\nu)} &=& -{G_F^2 \over 32 \pi^2} {1 \over E_\nu^2} 
\int_0^\infty q dq \int_{-k}^{\omega_{\rm max}} d \omega\ {\left(1 -f (E_\nu') 
\right) \over 1-e^{-z} }\ {\cal I}m \left(L^{\alpha\beta} \Pi^R_{\alpha\beta} 
\right) 
\label{mfp} \\
&& {\rm with} \quad \omega_{\rm \max}=\min(2 E_\nu -k,k) \nonumber
\eeq
$(1 -f (E_\nu')=\left( 1+\exp[(E_\nu'-\mu_\nu)/T])\right)$ is a Pauli
blocking factor for the outgoing lepton. The chemical potential 
$\mu_\nu$ is determined by $\beta$ equilibrium and by neutrino 
transport equations ({\it e.g.}, diffusion equation), which
determine the lepton fraction $Y_L$ at a given instant of 
the protoneutron star cooling.

In the following figure (Fig. 8), we represented the ratio of 
total cross sections as calculated in the RPA and mean 
field approximations. The chosen thermodynamical conditions
are representative of the earlier stage of the cooling, 
when the neutrinos are still trapped inside the matter
(the lefpton fraction was chosen to be $Y_L$=0.4, for 
typical values of the temperature ($T=20$ MeV) and 
density ($n=4\, n_{\rm sat}$). The neutrino energy 
was fixed to $E_\nu=30$ MeV. 

\begin{figure}[htb]
\mbox{%
\parbox{10cm}{\epsfig{file=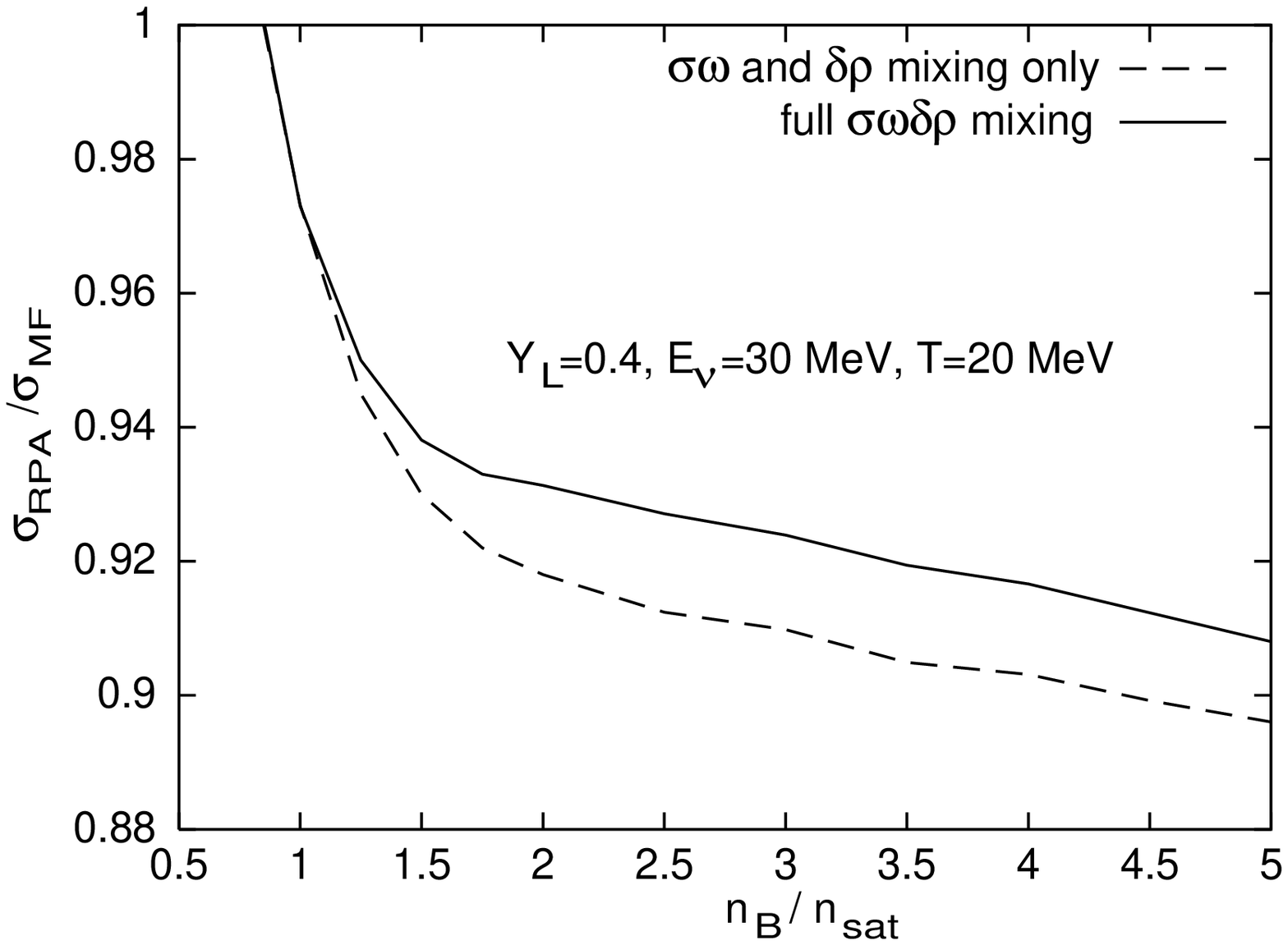,width=10cm}}
\parbox{5.6cm}{{\bf Fig. 8} -- Reduction factor in matter with 
trapped neutrinos}
}
\end{figure} 

At high density, the total neutrino-neutron scattering 
cross section is found to be reduced by RPA correlations 
by a factor (only) $\sim  10 \% $. When full meson mixing 
is taken into account, the reduction is even smaller. 
This value should be compared to suppression factors
of the order of two quoted in recent publications 
\cite{RPLP99,YT00}. A tentative explanation for this 
discrepancy was given in section \S \ref{prelim}.
Further investigation on this issue is needed.

\begin{figure}[htb]
\mbox{%
\parbox{7.2cm}{\epsfig{file=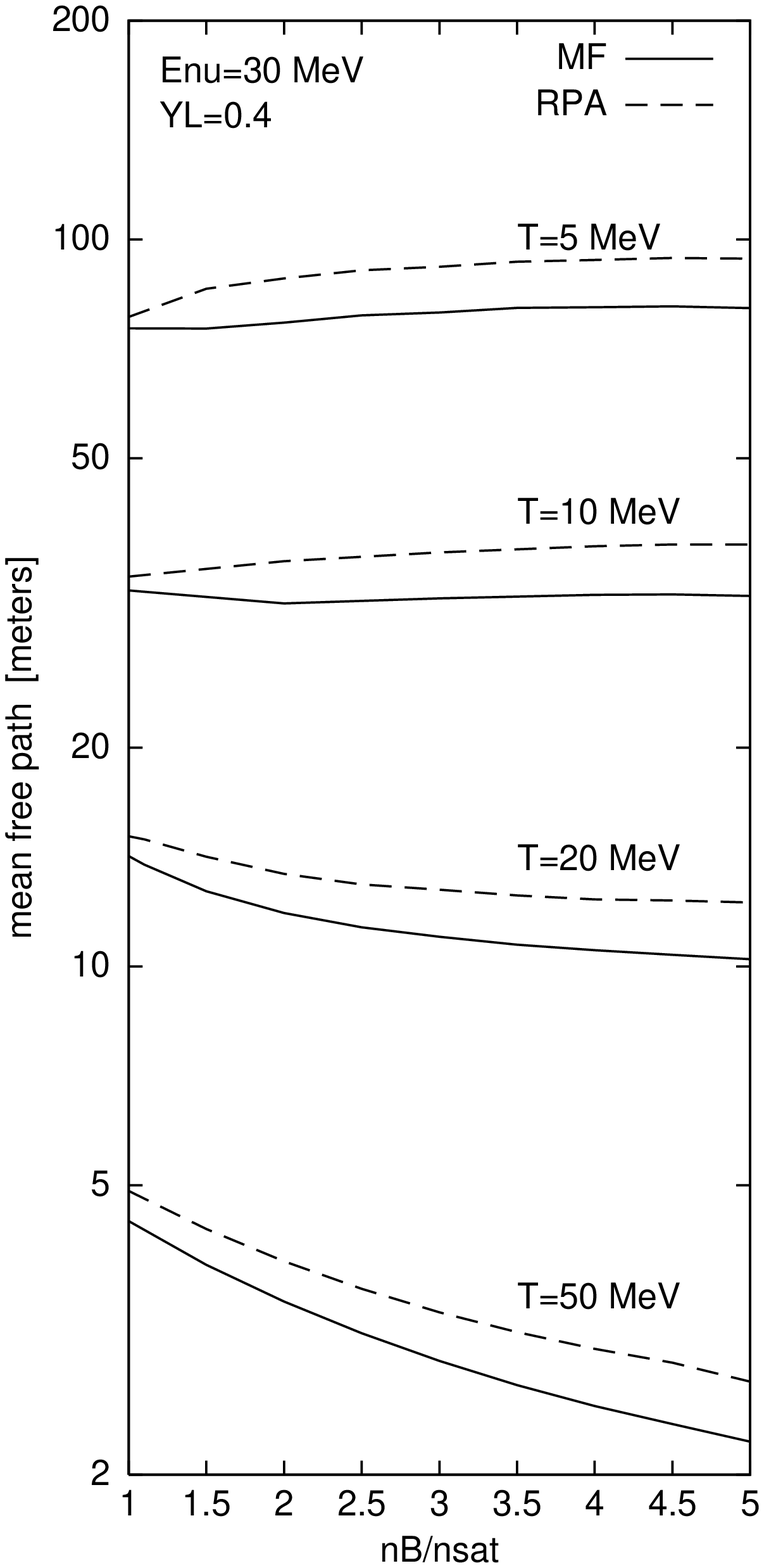,width=7.2cm}}
\parbox{7.2cm}{{\bf Fig. 9} -- 
Mean free path in matter with trapped neutrinos}
}
\end{figure}

The mean free path obtained from the integration of 
(\ref{mfp}) is shown on Fig. 9 in matter with trapped neutrinos
as a function of density for various values of the temperature. 
At high density, the mean free path is somewhat lenghtened 
by RPA correlations as compared to the mean field value.
At low density and moderate temperature, on the other hand, 
RPA correlations would yield an enhancement of the cross section 
and a reduction of the mean free path. A similar behavior 
was reported in \cite{YT00}. It should be kept in mind, however, 
that the validity of the model becomes questionable in this range.

\section{Summary and perspectives}

Let us gather here the main results obtained in this work.
Special attention was paid to the issues of renormalization of the
Dirac sea, residual interactions in the tensor channel and meson 
mixing. It has been shown that the vacuum fluctuations have only 
a negligible effect on the scattering rate. We examined two different 
prescriptions to introduce the contact term in the tensor channel,
and argue that, when properly taken into account, the numerical 
value of the 
Landau-Migdal parameter does not affect the structure functions.
As a consequence, the transverse contribution to the scattering rate 
is basically unchanged, while the contributions coming from the 
longitudinal and axial-vector parts combine to produce a 10\% to 
15 \% reduction of the scattering rate with respect to the mean 
field approximation. 

A natural extension of this work concerns the absorption and
emission cross section by the charged current reaction in the
RPA approximation. This is presently under study.

As the density increases, a more complete description should include 
the hyperons $\Lambda$, $\Sigma^-$. Taking into account 
these baryons does not present any particular difficulty, since the
structure of the equations is the same, only with different values
of the coupling constants. This has been done in \cite{RPL98,PRPLM99}.
As these authors argue moreover, the high fraction of trapped
neutrinos and relatively hot temperatures both conspire to
strongly suppress the formation of exotic, strangeness carrying
particles, so that these are not expected to have a strong influence 
on the first few seconds of the life of the protoneutron star.
Nevertheless, the exoticas will be determinant on the later
stage of the protoneutron star cooling, at $t \simeq 30-60\ s$, which 
are marginally or will soon become observable with the next 
generation of neutrino detectors now on construction, such as UNO.

A more important problem is to arrive to an accurate description 
of the short range correlations. These have been studied separately
from RPA corrections by several autors (see {\it eg. } 
\cite{RSS96,RS95,Sedra,Voskr}). The calculated correction factors 
to the neutrino opacities are large, especially in the low density
regime. This is in fact related to the well-known problem of
nuclear matter descriptions, which should simultaneously take
into account short and long range corrections, ``ladders'' and 
``loops'', in a consistent way. While methods do exist to do so
\cite{laddersnloops}, at least in an approximate way, 
they are quite unwieldy, especially in the relativistic formulation. 
Some results in this direction have been reported recently in \cite{Sedra}.

In matter at subnuclear density, and also at high density,
the homogeneous phase may not be the most stable state of matter.
It will therefore be important to study the role of ordered 
configurations, as coherent scattering is expected to dominate 
in this regime. At low density, and sufficiently low temperatures, 
the spinodal instasbility triggers condensation of droplets of 
a denser phase in a more dilute gas principally composed of neutrons.
The droplets take a spatially ordered configuration due to Coulomb
forces. This will eventually form the crust of the cooled neutron 
star. At high density, there are various possibilities of forming
ordered structures. For example, if a transition to quark gluon 
plasma takes place, an important fraction of the matter of star 
can be in a mixed phase, and will acquire an ordered structure 
similar to that present in the crust \cite{Glend}. Another 
possibility is the formation of a pion condensate in the alternating 
spin layer configuration \cite{ALS}. In any case, such structures 
are generally unfavored by temperature and high lepton (and proton) 
fractions, so that the same remark as done before applies: 
this type of correction are expected to set in at a later phase
of the protoneutron star cooling. It could be very interesting once
we are able to detect the tail of the neutrino emission from a 
supernova event; if a sudden change in the neutrino would occur, 
it could be interpreted as a signal that a phase transition has 
taken place.

\vskip 1cm

{\Large{\bf Acknowledgements}}

This work was supprted in part by  the Spanish Grants 
n$^{\underline{\rm o}}$ MCT-00-BFM-0357, DGES PB97-1432 
and AEN 99-0692.

\end{document}